\documentclass[aps,prl,a4paper,10pt,twocolumn,showpacs,floatfix,superscriptaddress,amsmath,amsfonts,amssymb,preprintnumbers]{revtex4-1}
\usepackage{float}

\usepackage{dcolumn,graphicx,color,booktabs,microtype,afterpage}

\graphicspath{{./}{figure/}}
\usepackage[charter,greekuppercase=italicized]{mathdesign}
\usepackage{sidecap}
\renewcommand{\tablename}{Table}
\makeatletter\renewcommand{\fnum@figure}[1]{\figurename~\thefigure.~}\makeatother
\makeatletter\renewcommand{\fnum@table}[1]{\tablename~\thetable.}\makeatother

\newcount\hh \newcount\mm
\hh=\time \divide\hh by 60
\mm=\hh \multiply\mm by 60 \mm=-\mm
\advance\mm by \time
\def\now{\number\hh:\ifnum\mm<10{}0\fi\number\mm}

\usepackage[colorlinks,plainpages=false,linkcolor=blue,urlcolor=blue,citecolor=blue,pdfpagemode=UseNone,pdfstartview=FitBH]{hyperref}

\newcommand{\tcr}[1]{\textcolor{black}{#1}}
\newcommand{\tcb}[1]{\textcolor{blue}{#1}}

\begin{document}

\makeatletter\renewcommand{\ps@plain}{%
\def\@evenhead{\hfill\itshape\rightmark}%
\def\@oddhead{\itshape\leftmark\hfill}%
\renewcommand{\@evenfoot}{\hfill\small{--~\thepage~--}\hfill}%
\renewcommand{\@oddfoot}{\hfill\small{--~\thepage~--}\hfill}%
}\makeatother\pagestyle{plain}

\preprint{\textit{Preprint: \today, \now. }}

\title{Simultaneous Nodal Superconductivity and Time-Reversal Symmetry Breaking in the Noncentrosymmetric Superconductor CaPtAs}
\author{T.\ Shang}\email[Corresponding authors:\\]{tian.shang@psi.ch}
\affiliation{Laboratory for Multiscale Materials Experiments, Paul Scherrer Institut, Villigen CH-5232, Switzerland}

\affiliation{Physik-Institut, Universität Z\"{u}rich, Winterthurerstrasse 190, CH-8057 Z\"{u}rich, Switzerland}
%
\author{M.\ Smidman}
\affiliation{Center for Correlated Matter and Department of Physics, Zhejiang University, Hangzhou 310058, China}
\author{A.\ Wang}
\affiliation{Center for Correlated Matter and Department of Physics, Zhejiang University, Hangzhou 310058, China}
\author{L.\ -J.\ Chang}
\affiliation{Department of Physics, National Cheng Kung University, Tainan 70101, Taiwan}
\author{C.\ Baines}
\affiliation{Laboratory for Muon-Spin Spectroscopy, Paul Scherrer Institut, CH-5232 Villigen PSI, Switzerland}
\author{M.\ K.\ Lee}
\affiliation{Department of Physics, National Cheng Kung University, Tainan 70101, Taiwan}
\author{Z.\ Y.\ Nie}
\affiliation{Center for Correlated Matter and Department of Physics, Zhejiang University, Hangzhou 310058, China}
\author{G.\ M.\ Pang}
\affiliation{Center for Correlated Matter and Department of Physics, Zhejiang University, Hangzhou 310058, China}
\author{W.\ Xie}
\affiliation{Center for Correlated Matter and Department of Physics, Zhejiang University, Hangzhou 310058, China}
\author{W.\ B.\ Jiang}
\affiliation{Center for Correlated Matter and Department of Physics, Zhejiang University, Hangzhou 310058, China}
\author{M.\ Shi}
\affiliation{Swiss Light Source, Paul Scherrer Institut, Villigen CH-5232, Switzerland}
\author{M.\ Medarde}
\affiliation{Laboratory for Multiscale Materials Experiments, Paul Scherrer Institut, Villigen CH-5232, Switzerland}
\author{T.\ Shiroka}
\affiliation{Laboratory for Muon-Spin Spectroscopy, Paul Scherrer Institut, CH-5232 Villigen PSI, Switzerland}
\affiliation{Laboratorium f\"ur Festk\"orperphysik, ETH Z\"urich, CH-8093 Zurich, Switzerland}
\author{H.\ Q.\ Yuan}\email{hqyuan@zju.edu.cn}
\affiliation{Center for Correlated Matter and Department of Physics, Zhejiang University, Hangzhou 310058, China}
\affiliation{Collaborative Innovation Center of Advanced Microstructures, Nanjing Univeristy, Nanjing 210093, China}

\begin{abstract}
By employing a series of experimental techniques,  
we provide clear evidence that CaPtAs represents a rare example of a noncentrosymmetric 
superconductor which simultaneously exhibits \emph{nodes} in the superconducting gap 
and \emph{broken time-reversal symmetry} (TRS) in its superconducting state 
(below $T_c$ $\approx$ 1.5\,K).
Unlike in fully-gapped superconductors, the magnetic penetration depth 
$\lambda(T)$ does not saturate at low temperatures, but instead it shows a 
$T^2$-dependence, characteristic of gap nodes.
Both the superfluid density and the electronic specific heat are 
best described by a two-gap 
model comprising of a nodeless gap and a gap with nodes, rather than by
single-band  
models. At the same time, zero-field muon-spin spectra exhibit increased 
relaxation rates below the onset of superconductivity, implying 
that TRS is broken in the superconducting state of CaPtAs, 
hence indicating its unconventional nature. 
Our observations suggest  CaPtAs to be \tcr{a \emph{new remarkable material} which links} 
two apparently 
disparate classes, that of TRS-breaking correlated magnetic superconductors with nodal gaps 
and the weakly-correlated noncentrosymmetric superconductors 
with broken TRS, normally exhibiting only a fully-gapped behavior.
\end{abstract}



\maketitle\enlargethispage{3pt}

\vspace{-5pt}

When entering the superconducting state, the breaking of extra 
symmetries in addition to $U(1)$ gauge symmetry is normally an 
indication of unconventional superconductivity (SC)~\cite{Sigrist1991,Tsuei2000}. 
In a growing number of superconductors, time-reversal symmetry (TRS) 
breaking has been proved 
via the detection of spontaneous magnetic fields below the onset of 
superconductivity by means of zero-field muon-spin relaxation 
measurements. Notable examples include 
Sr$_2$RuO$_4$~\cite{Luke1998}, UPt$_3$~\cite{Luke1993}, PrOs$_4$Sb$_4$~\cite{aoki2003}, LaNiGa$_2$~\cite{Hillier2012}, LaNiC$_2$, La$_7$$T_3$, and 
Re$T$ ($T$ = transition metal) superconductors~\cite{Hillier2009,Barker2015,singh2018La7Rh3,Singh2014,Shang2018,ShangReNb,Singh2017}. 
The first two are well-known examples of SC in 
strongly-correlated 
systems with unconventional pairing 
mechanisms~\cite{Mackenzie2003,Rober2002}, while the latter three are 
examples of noncentrosymmetric superconductors (NCSCs),
where the lack of inversion symmetry gives rise to an antisymmetric 
spin-orbit coupling (ASOC) leading to spin-split Fermi surfaces.
Consequently, their pairing states are not constrained to be purely singlet or triplet, 
and \emph{mixed-parity} pairing may 
occur~\cite{Bauer2012,Sungkit2014,smidman2017}. 
Owing to such mixed pairing and/or the influence of ASOC, NCSCs may 
exhibit significantly different properties from their conventional 
counterparts, e.g., superconducting gaps with 
nodes~\cite{yuan2006,nishiyama2007,bonalde2005CePt3Si,K2Cr3As3Pen,K2Cr3As3MuSR}, 
upper critical fields exceeding the Pauli limit~\cite{bauer2004,Carnicom2018,Shang2018,kimura2007,Chen2011} 
or, as recently proposed, even topological superconductivity~\cite{Kim2018,Sun2015,Ali2014,Sato2009,Tanaka2010}.

In general, the relationship between the breaking of TRS and a lack 
of inversion symmetry in the crystal structure is unclear. In many NCSCs 
such as Mo$_3$Al$_2$C, La$T$Si$_3$, Mg$_{10}$Ir$_{19}$B$_{16}$, or 
Mo$_3$P~\cite{bauer2010,Anand2011,Anand2014,Smidman2014,Acze2010,Shang2019}, 
no spontaneous magnetic fields have been observed and thus TRS is preserved 
in the superconducting state.
A notable feature of most of the weakly correlated NCSCs with broken TRS 
is the presence of fully opened superconducting gaps. 
\tcr{In the case of LaNiC$_2$, inconsistent results, including both fully-opened and nodal gap structures, have been found from measurements of the order parameter~\cite{Lee1996,chen2013,Hirose2012,chen2013,Bonalde2011,Landaeta2017}\footnote{In  an early report of the specific heat  of LaNiC$_2$, a $T^2$-dependence  of C/T was observed (indicating nodal SC)~\cite{Lee1996}, but more recently exponential behavior of $C/T$ was reported, consistent with a fully-gapped superconducting state~\cite{chen2013,Hirose2012}. Similar inconsistencies are also found from magnetic penetration depth $\lambda(T)$ measurements, 
where both a $T^2$- and an exponential temperature dependence have been reported~\cite{chen2013,Bonalde2011,Landaeta2017}, consistent with the presence of point nodes and fully gapped behavior, respectively.}.		
} 
The nodeless superconductivity of weakly correlated NCSCs is not only in contrast to the general expectations for strong singlet-triplet mixing, but also sets these systems apart from the strongly correlated superconductors Sr$_2$RuO$_4$ and UPt$_3$~\cite{Luke1998,Luke1993}, where the presence of unconventional pairing mechanisms is more unambiguously determined.

In LaNiC$_2$,  
as well as in centrosymmetric LaNiGa$_2$, the observed TRS breaking has been accounted for by non-unitary triplet pairing~\cite{Hillier2009, Quintanilla2010,Hillier2012}. This was reconciled with nodeless multigap SC by the proposal of even-parity triplet pairing, between electrons on different orbitals~\cite{Weng2016}.
On the other hand, the Re$T$ superconductors, which have a relatively 
large ASOC compared to LaNiC$_2$, appear to exhibit single fully-opened gaps, 
more consistent with a predominantly singlet pairing. 
The recent observation of TRS breaking in centrosymmetric elemental Re 
strongly suggests that the local electronic structure of Re is crucial 
for understanding the TRS breaking in 
the Re$T$ family~\cite{ShangReNb}.  
The broken TRS in weakly correlated systems, which otherwise appear to 
behave as
conventional superconductors, has led to proposals to account 
for this behavior with a conventional pairing mechanism~\cite{Agterberg1999}, such as 
the loop-Josephson-current state, based on a model with onsite singlet pairing~\cite{Ghosh2018}.

To date, there are \tcr{scarcely any} examples of NCSC which clearly exhibit 
broken TRS and nodal-gap SC. 
In this Letter, we show that CaPtAs, 
a newly discovered NCSC~\cite{Xie2019},  
is a rare candidate to display 
both such unconventional features. 
Our key observations of a nodal-gap 
and of spontaneous magnetic fields 
(concomitant with the onset of SC)  
indicate that CaPtAs represents \tcr{a new remarkable} 
example of a weakly-correlated NCSC encompassing
both broken TRS and nodal SC. 

%
%

Polycrystalline CaPtAs was synthesized via a solid-state reaction method~\cite{Xie2019}. 
Magnetic susceptibility, electrical resistivity, and specific-heat measurements were performed on 
a Quantum Design MPMS and PPMS, respectively. 
The  muon-spin relaxation/rotation ($\mu$SR) measurements were carried out on the 
low-temperature facility (LTF) spectrometers of the $\pi$M3 beamline at the Paul Scherrer Institut, Villigen, Switzerland. The temperature-dependent shift of the magnetic penetration depth, which is proportional to the frequency shift, i.e., $\Delta \lambda = G \Delta f$ (with $G$ a geometry related constant), 
was measured by using a tunnel-diode oscillator (TDO) based technique at an operating frequency of 7\,MHz~\cite{chen2013,K2Cr3As3Pen,Chen2011}.
%
%
%

\begin{figure}[tb]
	\centering
	\includegraphics[width=0.47\textwidth]{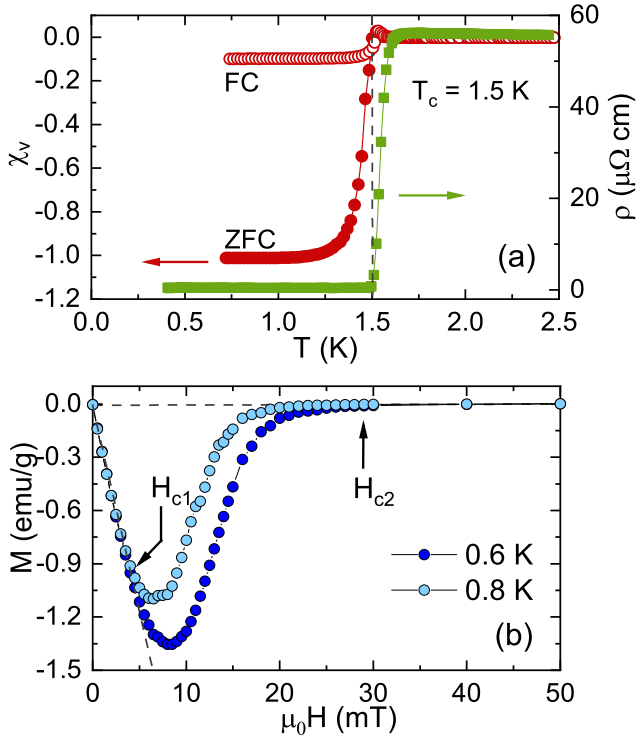}
	\caption{\label{fig:superconductivity}(a) Temperature dependent ZFC- and FC magnetic susceptibilities of CaPtAs (left axis), measured in an applied field 
	of 1\,mT, and zero-field electrical resistivity (right axis). \tcr{The magnetic susceptibility data were corrected after considering the demagnetization factor.} (b) Magnetization vs.\ applied magnetic field in the superconducting state. 
	The lower critical field $\mu_0H_\mathrm{c1}$ 
	was determined as the value where $M(H)$ starts deviating from linearity (see dashed-line), 
	while the upper critical field $\mu_0H_\mathrm{c2}$ was  
	identified with	the  field where the diamagnetic signal disappears.}
\end{figure}
%

CaPtAs crystallizes in a tetragonal noncentrosymmetric structure with space group  $I4_1md$ (No.~109)~\cite{Xie2019}. 
The 
SC of CaPtAs was characterized by magnetic susceptibility, 
measured using both field-cooling- (FC) and zero-field-cooling (ZFC) protocols. As shown in Fig.~\ref{fig:superconductivity}(a), the ZFC-susceptibility (\tcr{after accounting for the demagnetization factor}) indicates SC below $T_c = 1.5$\,K, where the electrical resistivity 
(right axis) drops to zero,  both being
consistent with the specific-heat data~\cite{Supple}.  
The lower-critical-field, estimated from the field-dependent magnetization, 
is $\mu_{0}H_{c1} = 4.8(1)$\,mT [see Fig.~\ref{fig:superconductivity}(b)].

%
\begin{figure}[tbh]
	\centering
	\includegraphics[width=0.48\textwidth]{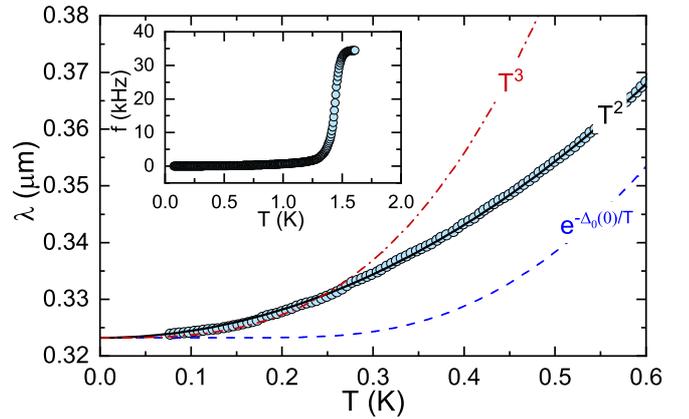}
	\caption{\label{fig:TDO} The magnetic penetration depth  $\lambda(T)$ of CaPtAs, 
	measured using the TDO-based method in zero field at $T \lesssim 1/3T_c$. 
	The inset shows the TDO frequency up to 
	temperatures above $T_c$. The black- and red lines represent fits to 
     $\lambda(T) \sim$ $T^\mathrm{2}$ and $T^\mathrm{3}$, respectively, while the blue line indicates 
     an exponential temperature dependence. $\lambda$ was calculated as $\lambda_0 + \Delta \lambda$, where $\lambda_0$ was derived from TF-$\mu$SR measurements.} 
\end{figure}
%

Figure~\ref{fig:TDO} shows the temperature dependent magnetic penetration depth $\lambda(T)$ measured by the TDO method 
and the corresponding exponential- and power-law fits.
The TDO data over the full temperature range (see inset) illustrate the superconducting transition near $T_c = 1.5$\,K. Clearly, $\lambda(T)$ follows a quadratic temperature dependence [$\lambda(T)$ $\sim$ $T^\mathrm{2}$], as expected for superconductors with point nodes. 
In contrast, a power-law with a larger exponent [$\lambda(T)$ $\sim$ $T^\mathrm{3}$] or an exponential temperature dependence [$\lambda(T)$ $\sim$ $e^{-\Delta_0(0)/T}$], the latter indicating fully-gapped behaviour,  both deviate significantly from the experimental data.     
 
%
\begin{figure}[!tbh]
	\centering
	\includegraphics[width=0.48\textwidth,angle=0]{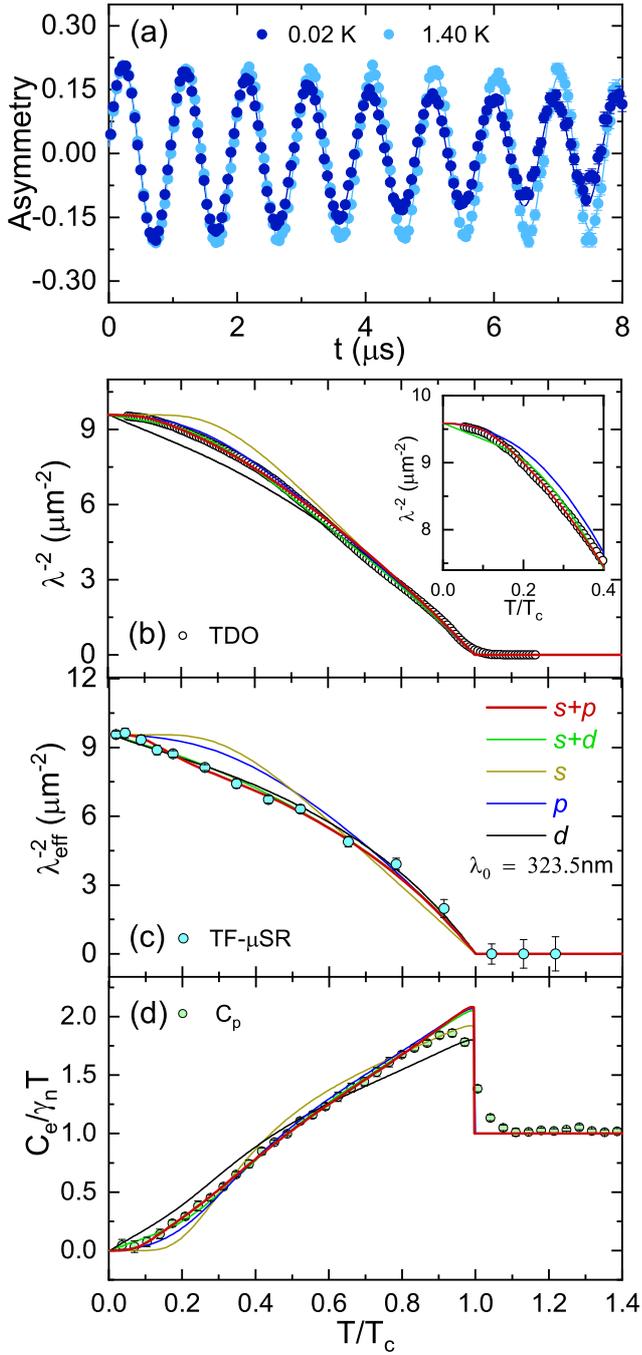}
	\vspace{-2ex}%
	\caption{\label{fig:TF_MuSR}(a) Time-domain TF-$\mu$SR spectra in 
	the superconducting (0.02\,K) and normal (1.4\,K) states of CaPtAs. 
	Superfluid density as estimated from (b) The TDO-based method and (c)
	TF-$\mu$SR vs.\ the reduced temperature $T$/$T_c$. The inset
 shows the enlarged plot of the TDO low-$T$ region. 
	(d) Zero-field electronic specific heat 
	vs.\ $T$/$T_c$. The different lines represent fits to the various models (see text for details).  
	The fit parameters are listed in Table~SI~\cite{Supple}.} 
\end{figure}
%

Figure~\ref{fig:TF_MuSR}(a) shows two typical 
transverse-field (TF) $\mu$SR 
spectra collected at temperatures above and below $T_c$ at 8\,mT ($T_c^\mathrm{8\,mT}$ = 1.15\,K)~\cite{Supple},
here corresponding to  
nearly twice $\mu_{0}H_{c1}(0)$. 
The TF-$\mu$SR asymmetry data were analyzed using: 
\begin{equation}
\label{eq:procession}
A_\mathrm{TF} = A_\mathrm{s}  e^{- \sigma^2 t^2/2} \cos(\gamma_{\mu} B_\mathrm{s} t + \phi) +
A_\mathrm{bg} \cos(\gamma_{\mu} B_\mathrm{bg} t + \phi).
\end{equation}
Here $A_\mathrm{s}$ (50\%) and $A_\mathrm{bg}$ (50\%) represent the asymmetry of the sample and background (e.g., sample holder), respectively.  
$\gamma_{\mu}/2\pi = 135.53$\,MHz/T is the muon gyromagnetic ratio, $B_\mathrm{s}$ and $B_\mathrm{bg}$ are the local fields sensed by implanted muons in the sample and sample holder,
$\phi$ is the shared initial phase, and $\sigma$ is a Gaussian relaxation rate. 
$\sigma$ includes contributions from both the  flux-line lattice ($\sigma_\mathrm{sc}$) and a  temperature-invariant relaxation due to nuclear moments ($\sigma_\mathrm{n}$).  
By subtracting the nuclear contribution in quadrature, one can extract 
$\sigma_\mathrm{sc}$, i.e., $\sigma_\mathrm{sc}$ = $\sqrt{\sigma^{2} - \sigma^{2}_\mathrm{n}}$.
The upper critical field 
of CaPtAs is relatively small compared to the field applied during the 
TF-$\mu$SR measurements ($H_{c2}$/$H_\mathrm{appl}$ $\sim$ 4.3)
~\cite{tinkham1996,Werthamer1966,Supple}. 
Hence, the effective penetration depth $\lambda_\mathrm{eff}$ had to be  
calculated from $\sigma_\mathrm{sc}$ by considering  
the overlap of vortex cores~\cite{Brandt2003}:
\begin{equation}
\label{eq:sigma_to_lambda}
\sigma_\mathrm{sc} (h) = 0.172 \frac{\gamma_{\mu} \Phi_0}{2\pi}(1-h)[1+1.21(1-\sqrt{h})^3]\lambda^{-2}_\mathrm{eff}.
\end{equation}
%
%
Here, $h = H_\mathrm{appl}/H_\mathrm{c2}$, is the reduced magnetic field. 

Figures~\ref{fig:TF_MuSR}(b)-(c) show the superfluid density 
($\rho_{\mathrm{sc}} \propto \lambda^{-2}$) measured by $\mu$SR and 
TDO vs.\ the reduced temperature $T/T_c$, respectively. 
The superfluid density clearly varies with temperature 
down to the lowest $T$, i.e., well below $T/T_c = 0.3$.  
This non-constant behavior again indicates the presence of low energy excitations and, hence, of nodes 
in the superconducting gap. To get further insight into the pairing symmetry, the temperature-dependent superfluid density was analyzed 
using different models. 
Considering a superconducting gap $\Delta_\mathrm{k}$, the superfluid density $\rho_\mathrm{sc}(T)$ can be calculated as:
\begin{equation}
\label{eq:rhos}
\rho_\mathrm{sc} = 1 + 2 \Bigg{\langle} \int^{\infty}_{\Delta_\mathrm{k}} \frac{E}{\sqrt{E^2-\Delta_\mathrm{k}^2}} \frac{\partial f}{\partial E} dE \Bigg{\rangle}_\mathrm{FS},
\end{equation}
where $f = (1+e^{E/k_\mathrm{B}T})^{-1}$ is the Fermi function and $\langle \rangle_\mathrm{FS}$ represents an average over the Fermi surface. The gap function can be written as $\Delta_\mathrm{k}(T) = \Delta_0(T) g_\mathrm{k}$, where $\Delta_0$ is the maximum gap value and $g_\mathrm{k}$ is the angular dependence of the gap (see details 
in Table~SI) \cite{Supple}. The temperature dependence of the gap was assumed to follow 
$\Delta_0(T) = \Delta_0(0) \mathrm{tanh} \{ 1.82[1.018(T_\mathrm{c}/T-1)]^{0.51} \}$, where
$\Delta_0(0)$ is the gap value in the zero-temperature limit.

Five different models, single-gap $s$-, $p$-, $d$-, and two-gap $s+p$- and $s+d$-wave, 
were used to analyze the superfluid density. 
The marked temperature dependence of the superfluid density at low-$T$  
clearly rules out a fully-gapped $s$-wave model [see yellow lines in Figs.~\ref{fig:TF_MuSR}(b)-(c)]. Also in the
case of a pure $p$-wave,  
we find a poor agreement with the low-$T$ data (blue lines). 
A $d$-wave model with line nodes, can reproduce reasonably well the 
TF-$\mu$SR data, but it fails to follow the low-$T$ $\lambda^{-2}(T)$ 
data obtained via TDO [see black lines in Figs.~\ref{fig:TF_MuSR}(b)-(c)].  
The slight difference between the TDO and TF-$\mu$SR data below 
$T/T_c \sim 0.5$ is most likely due to the applied external field (8\,mT) during the TF-$\mu$SR
measurements, which is not neglible comparedto the small $H_\mathrm{c2}$ value of CaPtAs.

Conversely, the superfluid density is best fitted by a 
two\--com\-po\-nent $s+p$- or $s+d$-wave model [red and green lines in Figs.~\ref{fig:TF_MuSR}(b)-(c)].
The good agreement with data of these  
models indicates the presence of multiple 
gaps, of which at least one has nodes on the Fermi surface. 
Although both models fit the superfluid density satisfactorily 
well across the full temperature range ($T < T_c$), the $s+p$-wave model agrees
better with the $\lambda^{-2}(T)$ data measured using the TDO method [see inset of Fig.~\ref{fig:TF_MuSR}(b)]. 
This is also strongly evidenced by both its smaller deviation from 
the data (see Table~SI) and the quadratic low-$T$ dependence of $\lambda(T)$~\cite{Supple}.   

To further validate the above conclusions, the zero-field electronic 
specific heat $C_\mathrm{e}/T$ was also analyzed using the above models~\cite{tinkham1996,Padamsee1973,Bouquet2001}. 
$C_\mathrm{e}/T$ was obtained by subtracting the phonon- and nuclear 
contributions from the measured data (see Fig.~S2)~\cite{Supple} and it is shown in Fig.~\ref{fig:TF_MuSR}(d) as $C_\mathrm{e}/\gamma_\mathrm{n}T$, 
with $\gamma_\mathrm{n}$ the normal-state electronic specific-heat coefficient.
Again, the single-gap $s$-, $p$-, and $d$-wave models deviate significantly from the data. 
Conversely, both multigap models exhibit a good agreement with the experimental data across the full temperature range, with the $s+p$-wave model showing 
the smallest deviation~\cite{Supple}, hence  
providing further evidence for nodal-gap SC in CaPtAs. 
The fit of the $s+p$-wave model to the superfluid density and the electronic specific heat  
[including the $s$-($\sim$15\%) and $p$-wave ($\sim$85\%) components] is shown in Fig.~S3~\cite{Supple}.

\begin{figure}[thb!]
	\includegraphics[width=0.45\textwidth,angle=0]{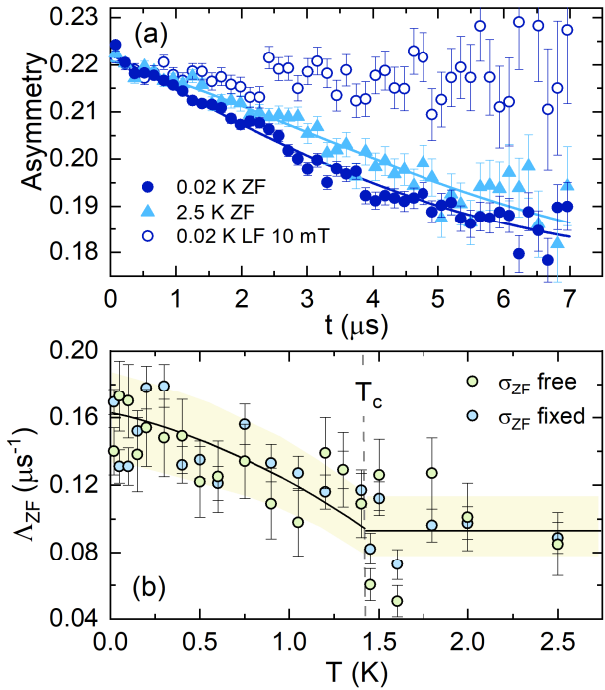}
	\vspace{-2ex}%
	\caption{\label{fig:ZF_muSR} (a) Representative zero-field $\mu$SR time spectra collected in the superconducting- (0.02\,K) and normal (2.5\,K) state of CaPtAs. Additional data were collected at 0.02\,K in a 10-mT longitudinal field.
		The solid lines are fits to Eq.~(\ref{eq:KT_and_electr}). (b) Derived Lorentzian relaxation rate $\Lambda_\mathrm{ZF}$ (using either a free- or a fixed-$\sigma_\mathrm{ZF}$ analysis) versus temperature. The solid line through the data is a guide to the eyes.} 
\end{figure}
%

To search for spontaneous fields below $T_c$, signaling possible TRS breaking in CaPtAs, we performed zero field (ZF)-$\mu$SR measurements. 
The clear increase in relaxation rate 
in the superconducting state [see Fig.~\ref{fig:ZF_muSR}(a)] hints at the breaking of TRS.  
For non-magnetic materials, the depolarization
is generally described by a Gaussian Kubo-Toyabe relaxation function~\cite{Kubo1967,Yaouanc2011}. 
For CaPtAs, the ZF-$\mu$SR spectra were fitted by considering an 
additional Lorentzian relaxation component, with $A_\mathrm{s}$ and $A_\mathrm{bg}$ 
being the same as in the TF-$\mu$SR case:
\begin{equation}
\label{eq:KT_and_electr}
A_\mathrm{ZF} = A_\mathrm{s}\left[\frac{1}{3} + \frac{2}{3}(1 -
\sigma_\mathrm{ZF}^{2}t^{2})\,
\mathrm{e}^{(-\frac{\sigma_\mathrm{ZF}^{2}t^{2}}{2})}\right]\mathrm{e}^{-\Lambda_\mathrm{ZF} t} + A_\mathrm{bg}.
\end{equation}
Fits using the above model yield an almost temperature-independent Gaussian relaxation rate ($\sigma_\mathrm{ZF}$) across the measured temperature range [see Fig.~S4(b)]~\cite{Supple}.  
Hence, the Lorentzian relaxation rate ($\Lambda_\mathrm{ZF}$) was
estimated by fixing $\sigma_\mathrm{ZF}$ to its average value ($\sigma_\mathrm{ZF}^\mathrm{avg}$ = 0.13~$\mu$s$^{-1}$). As shown in Fig.~\ref{fig:ZF_muSR}(b), a small yet measurable increase
of $\Lambda_\mathrm{ZF}$ below $T_c$ and a temperature-independent 
relaxation above $T_c$, reflect the onset of spontaneous magnetic fields. 
The latter can be considered as the signature of TRS breaking in the superconducting 
state of CaPtAs, with similarly enhanced $\Lambda_\mathrm{ZF}$ having also been 
found in other TRS breaking NCSCs~\cite{Hillier2009,Barker2015}. 
Both free- and fixed-$\sigma_\mathrm{ZF}$ analyses 
show a robust increase in $\Lambda_\mathrm{ZF}(T)$ below $T_c$, 
demonstrating that the signal of spontaneous magnetic fields 
is an intrinsic effect, rather than an artifact of correlated 
fit parameters. This is further confirmed in Fig.~S5, where we show the cross correlations
between the different fit parameters~\cite{Supple}. 
Finally, longitudinal-field (LF)-$\mu$SR measurements were  
performed at base temperature (0.02\,K) to rule out additional extrinsic effects such as defect/impurity induced relaxation. As shown in Fig.~\ref{fig:ZF_muSR}(a), a small field of 10\,mT is sufficient to fully decouple the muon spins from the weak spontaneous magnetic fields, indicating that the fields are static on the time scale of the muon lifetime.


To date, most NCSCs with broken TRS exhibit nodeless 
superconductivity, indicating that the spin-singlet channel dominates the  pairing. These include the $\alpha$-Mn-type Re$T$~\cite{Singh2014,Singh2017,Shang2018,ShangReNb}
and Th$_7$Fe$_3$-type La$_7$$T_3$~\cite{Barker2015,singh2018La7Rh3}. As for CeNiC$_2$-type NCSCs, the recently discovered ThCoC$_2$ exhibits 
nodal SC, but no evidence of broken TRS has been found~\cite{Bhattacharyya2019}.
In LaNiC$_2$ instead, the low symmetry of its orthorhombic crystal structure means that the breaking of TRS at $T_c$ necessarily implies nonunitary triplet 
pairing, \tcr{and rules out the mixed singlet-triplet state described below~\cite{Hillier2009,Quintanilla2010}. However, measurements of the gap symmetry have yielded inconsistent results, where both fully-opened and nodal gap structures have been reported~\cite{Lee1996,chen2013,Hirose2012,chen2013,Bonalde2011,Landaeta2017}.}
Compared to the above cases, CaPtAs \tcr{represents a new remarkable NSCS}, which 
accommodates both broken TRS and nodal SC.

\tcr{One possibility is that the observed multigap superconductivity corresponds to different gaps on distinct electronic bands, which would be consistent with band-structure calculations showing multiple bands crossing the Fermi level~\cite{Xie2019}. An alternative scenario is that} in NCSCs, the admixture of singlet- $\psi(\boldsymbol{k})$ and triplet $\boldsymbol{d}(\boldsymbol{k})$ order parameters leads to gap 
structures with 
$\Delta(\boldsymbol{k})_\pm = \psi(\boldsymbol{k}) \pm   \left| \boldsymbol{d}(\boldsymbol{k}) \right |$
~\cite{Bauer2012,Sungkit2014,smidman2017}. Clearly, if the triplet component is small, the gaps on the 
spin-split Fermi surfaces will both be nodeless and of nearly equal magnitude, making this case 
hardly distinguishable 
from a single-gap $s$-wave superconductor.
On the other hand, if the triplet component dominates, there can be one nodeless gap, and another with nodes. Such a scenario can explain well, for instance, the multigap nodeless SC in Li$_2$Pd$_3$B and the nodal SC
in Li$_2$Pt$_3$B, since the triplet-component increases with the  enhanced ASOC upon the substitution of Pt for Pd~\cite{yuan2006}. We find that the superfluid density of CaPtAs is best described by models with one nodeless and one nodal gap, which  also corresponds to that expected for significant singlet-triplet mixing. 

\tcr{According to band-structure calculations, the estimated band splitting due to ASOC is about 50--100\,meV, which gives $E_\mathrm{soc}$/$k_\mathrm{B}$$T_c$ $\sim$ 400-800~\cite{Xie2019}. Though 
much smaller than the band splitting of CePt$_3$Si ($E_\mathrm{soc}$/$k_\mathrm{B}$$T_c$ $\sim$ 3095)~\cite{Samokhin2004}, it is comparable to that of Li$_2$Pt$_3$B 
($\sim$ 831)~\cite{Lee2005}, and much larger than that of most  other NCSCs.
Since the above two Pt compounds are believed to exhibit mixed pairing~\cite{Frigeri2004,yuan2006}, 
the presence of large band splitting due to ASOC, in addition to nodal multigap SC, suggests that CaPtAs is a good candidate for large singlet-triplet mixing.}
However, whether in this crystal structure there is a TRS-breaking mixed singlet-triplet state compatible with the ASOC, requires further theoretical analysis. We note that, if one considers TRS breaking states corresponding to the two-dimensional irreducible representations of the point group $C_{4v}$, the simplest one is a chiral $p$-wave state~\cite{smidman2017}, 
originally applied to Sr$_2$RuO$_4$~\cite{Mackenzie2003}. 
The $p$-wave model and $p$-component of $s+p$-wave model
we use in Fig.~\ref{fig:TF_MuSR} correspond to this 
chiral $p$-wave state.
A specific attribution of the pairing symmetry requires further measurements on single crystals, together with microscopic 
calculations based on the band structure. Different from 
\tcb{the} proposed topological NCSCs~\cite{Kim2018,Sun2015}, 
CaPtAs exhibits simultaneously nodal SC and broken TRS. 
This suggests that it could be a possible exotic-type of topological superconductor, \tcr{a suitable candidate material in which to search for Majorana zero modes~\cite{Schnyder2015,Sato2017}.} 
Due to its high stability and the availability of 
single crystals~\cite{Xie2019}, CaPtAs is very promising for future investigations 
using other techniques, such as scanning tunneling microscopy (STM), or angle-resolved photoemission spectroscopy (ARPES).
 	
In summary, we find that CaPtAs is 
an example of an NCSC exhibiting both TRS breaking and nodal superconductivity. Its superfluid density and specific heat are best described by a two-gap model, with one gap being fully open and the other being nodal. 
\tcr{ The presence of multigap nodal superconductivity and sizeable band splitting due to ASOC makes CaPtAs a good candidate for hosting mixed}
singlet- and triplet pairing. While further theoretical calculations and  measurements are necessary to determine the nature of the order parameter and pairing mechanism, this system may offer new insights for bridging the gap between  different classes of TRS-breaking superconductors, namely strongly correlated superconductors with magnetically mediated pairing and nodal gaps 
(such as Sr$_2$RuO$_4$ and UPt$_3$) and the more recently discovered weakly-correlated NCSCs. 

%
This work was supported by the National Key R\&D Program of China
(Grants no.\ 2016YFA0300202 and 2017YFA0303100), the National
Natural Science Foundation of China (Grants no. U1632275, 11874320 and \tcr{11974306}), the Schwei\-ze\-rische Na\-ti\-o\-nal\-fonds zur 
F\"{o}r\-de\-rung der Wis\-sen\-schaft\-lich\-en For\-schung, SNF 
(Grants no.\ 200021-169455 and 206021-139082). L.\ J.\ C.\ thanks the 
MOST Funding for the support under the projects 104-2112-M-006-010-MY3 and 
107-2112-M-006-020. We also acknowledge the assistance from other beamline scientists on LTF $\mu$SR spectrometers at PSI.

\bibliography{CaPtAs_bib}

\begin{thebibliography}{63}%
\makeatletter
\providecommand \@ifxundefined [1]{%
 \@ifx{#1\undefined}
}%
\providecommand \@ifnum [1]{%
 \ifnum #1\expandafter \@firstoftwo
 \else \expandafter \@secondoftwo
 \fi
}%
\providecommand \@ifx [1]{%
 \ifx #1\expandafter \@firstoftwo
 \else \expandafter \@secondoftwo
 \fi
}%
\providecommand \natexlab [1]{#1}%
\providecommand \enquote  [1]{``#1''}%
\providecommand \bibnamefont  [1]{#1}%
\providecommand \bibfnamefont [1]{#1}%
\providecommand \citenamefont [1]{#1}%
\providecommand \href@noop [0]{\@secondoftwo}%
\providecommand \href [0]{\begingroup \@sanitize@url \@href}%
\providecommand \@href[1]{\@@startlink{#1}\@@href}%
\providecommand \@@href[1]{\endgroup#1\@@endlink}%
\providecommand \@sanitize@url [0]{\catcode `\\12\catcode `\$12\catcode
  `\&12\catcode `\#12\catcode `\^12\catcode `\_12\catcode `\%12\relax}%
\providecommand \@@startlink[1]{}%
\providecommand \@@endlink[0]{}%
\providecommand \url  [0]{\begingroup\@sanitize@url \@url }%
\providecommand \@url [1]{\endgroup\@href {#1}{\urlprefix }}%
\providecommand \urlprefix  [0]{URL }%
\providecommand \Eprint [0]{\href }%
\providecommand \doibase [0]{https://doi.org/}%
\providecommand \selectlanguage [0]{\@gobble}%
\providecommand \bibinfo  [0]{\@secondoftwo}%
\providecommand \bibfield  [0]{\@secondoftwo}%
\providecommand \translation [1]{[#1]}%
\providecommand \BibitemOpen [0]{}%
\providecommand \bibitemStop [0]{}%
\providecommand \bibitemNoStop [0]{.\EOS\space}%
\providecommand \EOS [0]{\spacefactor3000\relax}%
\providecommand \BibitemShut  [1]{\csname bibitem#1\endcsname}%
\let\auto@bib@innerbib\@empty
\bibitem [{\citenamefont {Sigrist}\ and\ \citenamefont
  {Ueda}(1991)}]{Sigrist1991}%
  \BibitemOpen
  \bibfield  {author} {\bibinfo {author} {\bibfnamefont {M.}~\bibnamefont
  {Sigrist}}\ and\ \bibinfo {author} {\bibfnamefont {K.}~\bibnamefont {Ueda}},\
  }\bibfield  {title} {\bibinfo {title} {Phenomenological theory of
  unconventional superconductivity},\ }\href
  {https://doi.org/10.1103/RevModPhys.63.239} {\bibfield  {journal} {\bibinfo
  {journal} {Rev. Mod. Phys.}\ }\textbf {\bibinfo {volume} {63}},\ \bibinfo
  {pages} {239} (\bibinfo {year} {1991})}\BibitemShut {NoStop}%
\bibitem [{\citenamefont {Tsuei}\ and\ \citenamefont
  {Kirtley}(2000)}]{Tsuei2000}%
  \BibitemOpen
  \bibfield  {author} {\bibinfo {author} {\bibfnamefont {C.~C.}\ \bibnamefont
  {Tsuei}}\ and\ \bibinfo {author} {\bibfnamefont {J.~R.}\ \bibnamefont
  {Kirtley}},\ }\bibfield  {title} {\bibinfo {title} {Pairing symmetry in
  cuprate superconductors},\ }\href {https://doi.org/10.1103/RevModPhys.72.969}
  {\bibfield  {journal} {\bibinfo  {journal} {Rev. Mod. Phys.}\ }\textbf
  {\bibinfo {volume} {72}},\ \bibinfo {pages} {969} (\bibinfo {year}
  {2000})}\BibitemShut {NoStop}%
\bibitem [{\citenamefont {Luke}\ \emph {et~al.}(1998)\citenamefont {Luke},
  \citenamefont {Fudamoto}, \citenamefont {Kojima}, \citenamefont {Larkin},
  \citenamefont {Merrin}, \citenamefont {Nachumi}, \citenamefont {Uemura},
  \citenamefont {Maeno}, \citenamefont {Mao}, \citenamefont {Mori},
  \citenamefont {Nakamura},\ and\ \citenamefont {Sigrist}}]{Luke1998}%
  \BibitemOpen
  \bibfield  {author} {\bibinfo {author} {\bibfnamefont {G.~M.}\ \bibnamefont
  {Luke}}, \bibinfo {author} {\bibfnamefont {Y.}~\bibnamefont {Fudamoto}},
  \bibinfo {author} {\bibfnamefont {K.~M.}\ \bibnamefont {Kojima}}, \bibinfo
  {author} {\bibfnamefont {M.~I.}\ \bibnamefont {Larkin}}, \bibinfo {author}
  {\bibfnamefont {J.}~\bibnamefont {Merrin}}, \bibinfo {author} {\bibfnamefont
  {B.}~\bibnamefont {Nachumi}}, \bibinfo {author} {\bibfnamefont {Y.~J.}\
  \bibnamefont {Uemura}}, \bibinfo {author} {\bibfnamefont {Y.}~\bibnamefont
  {Maeno}}, \bibinfo {author} {\bibfnamefont {Z.~Q.}\ \bibnamefont {Mao}},
  \bibinfo {author} {\bibfnamefont {Y.}~\bibnamefont {Mori}}, \bibinfo {author}
  {\bibfnamefont {H.}~\bibnamefont {Nakamura}},\ and\ \bibinfo {author}
  {\bibfnamefont {M.}~\bibnamefont {Sigrist}},\ }\bibfield  {title} {\bibinfo
  {title} {Time-reversal symmetry-breaking superconductivity in
  {Sr}$_{2}${RuO}$_{4}$},\ }\href {https://doi.org/10.1038/29038} {\bibfield
  {journal} {\bibinfo  {journal} {Nature}\ }\textbf {\bibinfo {volume} {394}},\
  \bibinfo {pages} {558} (\bibinfo {year} {1998})}\BibitemShut {NoStop}%
\bibitem [{\citenamefont {Luke}\ \emph {et~al.}(1993)\citenamefont {Luke},
  \citenamefont {Keren}, \citenamefont {Le}, \citenamefont {Wu}, \citenamefont
  {Uemura}, \citenamefont {Bonn}, \citenamefont {Taillefer},\ and\
  \citenamefont {Garrett}}]{Luke1993}%
  \BibitemOpen
  \bibfield  {author} {\bibinfo {author} {\bibfnamefont {G.~M.}\ \bibnamefont
  {Luke}}, \bibinfo {author} {\bibfnamefont {A.}~\bibnamefont {Keren}},
  \bibinfo {author} {\bibfnamefont {L.~P.}\ \bibnamefont {Le}}, \bibinfo
  {author} {\bibfnamefont {W.~D.}\ \bibnamefont {Wu}}, \bibinfo {author}
  {\bibfnamefont {Y.~J.}\ \bibnamefont {Uemura}}, \bibinfo {author}
  {\bibfnamefont {D.~A.}\ \bibnamefont {Bonn}}, \bibinfo {author}
  {\bibfnamefont {L.}~\bibnamefont {Taillefer}},\ and\ \bibinfo {author}
  {\bibfnamefont {J.~D.}\ \bibnamefont {Garrett}},\ }\bibfield  {title}
  {\bibinfo {title} {Muon spin relaxation in {U}{Pt}$_3$},\ }\href
  {https://doi.org/10.1103/PhysRevLett.71.1466} {\bibfield  {journal} {\bibinfo
   {journal} {Phys. Rev. Lett.}\ }\textbf {\bibinfo {volume} {71}},\ \bibinfo
  {pages} {1466} (\bibinfo {year} {1993})}\BibitemShut {NoStop}%
\bibitem [{\citenamefont {Aoki}\ \emph {et~al.}(2003)\citenamefont {Aoki},
  \citenamefont {Tsuchiya}, \citenamefont {Kanayama}, \citenamefont {Saha},
  \citenamefont {Sugawara}, \citenamefont {Sato}, \citenamefont {Higemoto},
  \citenamefont {Koda}, \citenamefont {Ohishi}, \citenamefont {Nishiyama},\
  and\ \citenamefont {Kadono}}]{aoki2003}%
  \BibitemOpen
  \bibfield  {author} {\bibinfo {author} {\bibfnamefont {Y.}~\bibnamefont
  {Aoki}}, \bibinfo {author} {\bibfnamefont {A.}~\bibnamefont {Tsuchiya}},
  \bibinfo {author} {\bibfnamefont {T.}~\bibnamefont {Kanayama}}, \bibinfo
  {author} {\bibfnamefont {S.~R.}\ \bibnamefont {Saha}}, \bibinfo {author}
  {\bibfnamefont {H.}~\bibnamefont {Sugawara}}, \bibinfo {author}
  {\bibfnamefont {H.}~\bibnamefont {Sato}}, \bibinfo {author} {\bibfnamefont
  {W.}~\bibnamefont {Higemoto}}, \bibinfo {author} {\bibfnamefont
  {A.}~\bibnamefont {Koda}}, \bibinfo {author} {\bibfnamefont {K.}~\bibnamefont
  {Ohishi}}, \bibinfo {author} {\bibfnamefont {K.}~\bibnamefont {Nishiyama}},\
  and\ \bibinfo {author} {\bibfnamefont {R.}~\bibnamefont {Kadono}},\
  }\bibfield  {title} {\bibinfo {title} {Time-reversal symmetry-breaking
  superconductivity in heavy-fermion {Pr}{Os}$_{4}${Sb}$_{12}$ detected by
  muon-spin relaxation},\ }\href
  {https://doi.org/10.1103/PhysRevLett.91.067003} {\bibfield  {journal}
  {\bibinfo  {journal} {Phys. Rev. Lett.}\ }\textbf {\bibinfo {volume} {91}},\
  \bibinfo {pages} {067003} (\bibinfo {year} {2003})}\BibitemShut {NoStop}%
\bibitem [{\citenamefont {Hillier}\ \emph {et~al.}(2012)\citenamefont
  {Hillier}, \citenamefont {Quintanilla}, \citenamefont {Mazidian},
  \citenamefont {Annett},\ and\ \citenamefont {Cywinski}}]{Hillier2012}%
  \BibitemOpen
  \bibfield  {author} {\bibinfo {author} {\bibfnamefont {A.~D.}\ \bibnamefont
  {Hillier}}, \bibinfo {author} {\bibfnamefont {J.}~\bibnamefont
  {Quintanilla}}, \bibinfo {author} {\bibfnamefont {B.}~\bibnamefont
  {Mazidian}}, \bibinfo {author} {\bibfnamefont {J.~F.}\ \bibnamefont
  {Annett}},\ and\ \bibinfo {author} {\bibfnamefont {R.}~\bibnamefont
  {Cywinski}},\ }\bibfield  {title} {\bibinfo {title} {Nonunitary triplet
  pairing in the centrosymmetric superconductor {La}{Ni}{Ga}$_2$},\ }\href
  {https://doi.org/10.1103/PhysRevLett.109.097001} {\bibfield  {journal}
  {\bibinfo  {journal} {Phys. Rev. Lett.}\ }\textbf {\bibinfo {volume} {109}},\
  \bibinfo {pages} {097001} (\bibinfo {year} {2012})}\BibitemShut {NoStop}%
\bibitem [{\citenamefont {Hillier}\ \emph {et~al.}(2009)\citenamefont
  {Hillier}, \citenamefont {Quintanilla},\ and\ \citenamefont
  {Cywinski}}]{Hillier2009}%
  \BibitemOpen
  \bibfield  {author} {\bibinfo {author} {\bibfnamefont {A.~D.}\ \bibnamefont
  {Hillier}}, \bibinfo {author} {\bibfnamefont {J.}~\bibnamefont
  {Quintanilla}},\ and\ \bibinfo {author} {\bibfnamefont {R.}~\bibnamefont
  {Cywinski}},\ }\bibfield  {title} {\bibinfo {title} {Evidence for
  time-reversal symmetry breaking in the noncentrosymmetric superconductor
  {LaNiC$_{2}$}},\ }\href {https://doi.org/10.1103/PhysRevLett.102.117007}
  {\bibfield  {journal} {\bibinfo  {journal} {Phys. Rev. Lett.}\ }\textbf
  {\bibinfo {volume} {102}},\ \bibinfo {pages} {117007} (\bibinfo {year}
  {2009})}\BibitemShut {NoStop}%
\bibitem [{\citenamefont {Barker}\ \emph {et~al.}(2015)\citenamefont {Barker},
  \citenamefont {Singh}, \citenamefont {Thamizhavel}, \citenamefont {Hillier},
  \citenamefont {Lees}, \citenamefont {Balakrishnan}, \citenamefont {Paul},\
  and\ \citenamefont {Singh}}]{Barker2015}%
  \BibitemOpen
  \bibfield  {author} {\bibinfo {author} {\bibfnamefont {J.~A.~T.}\
  \bibnamefont {Barker}}, \bibinfo {author} {\bibfnamefont {D.}~\bibnamefont
  {Singh}}, \bibinfo {author} {\bibfnamefont {A.}~\bibnamefont {Thamizhavel}},
  \bibinfo {author} {\bibfnamefont {A.~D.}\ \bibnamefont {Hillier}}, \bibinfo
  {author} {\bibfnamefont {M.~R.}\ \bibnamefont {Lees}}, \bibinfo {author}
  {\bibfnamefont {G.}~\bibnamefont {Balakrishnan}}, \bibinfo {author}
  {\bibfnamefont {D.~M.}\ \bibnamefont {Paul}},\ and\ \bibinfo {author}
  {\bibfnamefont {R.~P.}\ \bibnamefont {Singh}},\ }\bibfield  {title} {\bibinfo
  {title} {Unconventional superconductivity in {La}$_{7}${Ir}$_{3}$ revealed by
  muon spin relaxation: Introducing a new family of noncentrosymmetric
  superconductor that breaks time-reversal symmetry},\ }\href
  {https://doi.org/10.1103/PhysRevLett.115.267001} {\bibfield  {journal}
  {\bibinfo  {journal} {Phys. Rev. Lett.}\ }\textbf {\bibinfo {volume} {115}},\
  \bibinfo {pages} {267001} (\bibinfo {year} {2015})}\BibitemShut {NoStop}%
\bibitem [{\citenamefont {Singh}\ \emph {et~al.}(2018)\citenamefont {Singh},
  \citenamefont {Scheurer}, \citenamefont {Hillier},\ and\ \citenamefont
  {Singh}}]{singh2018La7Rh3}%
  \BibitemOpen
  \bibfield  {author} {\bibinfo {author} {\bibfnamefont {D.}~\bibnamefont
  {Singh}}, \bibinfo {author} {\bibfnamefont {M.~S.}\ \bibnamefont {Scheurer}},
  \bibinfo {author} {\bibfnamefont {A.~D.}\ \bibnamefont {Hillier}},\ and\
  \bibinfo {author} {\bibfnamefont {R.~P.}\ \bibnamefont {Singh}},\ }\bibfield
  {title} {\bibinfo {title} {Time-reversal-symmetry breaking and unconventional
  pairing in the noncentrosymmetric superconductor {La}$_{7}${Rh}$_{3}$ probed
  by $\mu${SR}},\ }\href@noop {} {\bibfield  {journal} {\bibinfo  {journal}
  {arXiv preprint arXiv:1802.01533}\ } (\bibinfo {year} {2018})}\BibitemShut
  {NoStop}%
\bibitem [{\citenamefont {Singh}\ \emph {et~al.}(2014)\citenamefont {Singh},
  \citenamefont {Hillier}, \citenamefont {Mazidian}, \citenamefont
  {Quintanilla}, \citenamefont {Annett}, \citenamefont {Paul}, \citenamefont
  {Balakrishnan},\ and\ \citenamefont {Lees}}]{Singh2014}%
  \BibitemOpen
  \bibfield  {author} {\bibinfo {author} {\bibfnamefont {R.~P.}\ \bibnamefont
  {Singh}}, \bibinfo {author} {\bibfnamefont {A.~D.}\ \bibnamefont {Hillier}},
  \bibinfo {author} {\bibfnamefont {B.}~\bibnamefont {Mazidian}}, \bibinfo
  {author} {\bibfnamefont {J.}~\bibnamefont {Quintanilla}}, \bibinfo {author}
  {\bibfnamefont {J.~F.}\ \bibnamefont {Annett}}, \bibinfo {author}
  {\bibfnamefont {D.~M.}\ \bibnamefont {Paul}}, \bibinfo {author}
  {\bibfnamefont {G.}~\bibnamefont {Balakrishnan}},\ and\ \bibinfo {author}
  {\bibfnamefont {M.~R.}\ \bibnamefont {Lees}},\ }\bibfield  {title} {\bibinfo
  {title} {Detection of time-reversal symmetry breaking in the
  noncentrosymmetric superconductor {Re}$_{6}${Zr} using muon-spin
  spectroscopy},\ }\href {https://doi.org/10.1103/PhysRevLett.112.107002}
  {\bibfield  {journal} {\bibinfo  {journal} {Phys. Rev. Lett.}\ }\textbf
  {\bibinfo {volume} {112}},\ \bibinfo {pages} {107002} (\bibinfo {year}
  {2014})}\BibitemShut {NoStop}%
\bibitem [{\citenamefont {Shang}\ \emph
  {et~al.}(2018{\natexlab{a}})\citenamefont {Shang}, \citenamefont {Pang},
  \citenamefont {Baines}, \citenamefont {Jiang}, \citenamefont {Xie},
  \citenamefont {Wang}, \citenamefont {Medarde}, \citenamefont {Pomjakushina},
  \citenamefont {Shi}, \citenamefont {Mesot}, \citenamefont {Yuan},\ and\
  \citenamefont {Shiroka}}]{Shang2018}%
  \BibitemOpen
  \bibfield  {author} {\bibinfo {author} {\bibfnamefont {T.}~\bibnamefont
  {Shang}}, \bibinfo {author} {\bibfnamefont {G.~M.}\ \bibnamefont {Pang}},
  \bibinfo {author} {\bibfnamefont {C.}~\bibnamefont {Baines}}, \bibinfo
  {author} {\bibfnamefont {W.~B.}\ \bibnamefont {Jiang}}, \bibinfo {author}
  {\bibfnamefont {W.}~\bibnamefont {Xie}}, \bibinfo {author} {\bibfnamefont
  {A.}~\bibnamefont {Wang}}, \bibinfo {author} {\bibfnamefont {M.}~\bibnamefont
  {Medarde}}, \bibinfo {author} {\bibfnamefont {E.}~\bibnamefont
  {Pomjakushina}}, \bibinfo {author} {\bibfnamefont {M.}~\bibnamefont {Shi}},
  \bibinfo {author} {\bibfnamefont {J.}~\bibnamefont {Mesot}}, \bibinfo
  {author} {\bibfnamefont {H.~Q.}\ \bibnamefont {Yuan}},\ and\ \bibinfo
  {author} {\bibfnamefont {T.}~\bibnamefont {Shiroka}},\ }\bibfield  {title}
  {\bibinfo {title} {Nodeless superconductivity and time-reversal symmetry
  breaking in the noncentrosymmetric superconductor {Re}$_{24}${Ti}$_5$},\
  }\href {https://doi.org/10.1103/PhysRevB.97.020502} {\bibfield  {journal}
  {\bibinfo  {journal} {Phys. Rev. B}\ }\textbf {\bibinfo {volume} {97}},\
  \bibinfo {pages} {020502(R)} (\bibinfo {year}
  {2018}{\natexlab{a}})}\BibitemShut {NoStop}%
\bibitem [{\citenamefont {Shang}\ \emph
  {et~al.}(2018{\natexlab{b}})\citenamefont {Shang}, \citenamefont {Smidman},
  \citenamefont {Ghosh}, \citenamefont {Baines}, \citenamefont {Chang},
  \citenamefont {Gawryluk}, \citenamefont {Barker}, \citenamefont {Singh},
  \citenamefont {Paul}, \citenamefont {Balakrishnan}, \citenamefont
  {Pomjakushina}, \citenamefont {Shi}, \citenamefont {Medarde}, \citenamefont
  {Hillier}, \citenamefont {Yuan}, \citenamefont {Quintanilla}, \citenamefont
  {Mesot},\ and\ \citenamefont {Shiroka}}]{ShangReNb}%
  \BibitemOpen
  \bibfield  {author} {\bibinfo {author} {\bibfnamefont {T.}~\bibnamefont
  {Shang}}, \bibinfo {author} {\bibfnamefont {M.}~\bibnamefont {Smidman}},
  \bibinfo {author} {\bibfnamefont {S.~K.}\ \bibnamefont {Ghosh}}, \bibinfo
  {author} {\bibfnamefont {C.}~\bibnamefont {Baines}}, \bibinfo {author}
  {\bibfnamefont {L.~J.}\ \bibnamefont {Chang}}, \bibinfo {author}
  {\bibfnamefont {D.~J.}\ \bibnamefont {Gawryluk}}, \bibinfo {author}
  {\bibfnamefont {J.~A.~T.}\ \bibnamefont {Barker}}, \bibinfo {author}
  {\bibfnamefont {R.~P.}\ \bibnamefont {Singh}}, \bibinfo {author}
  {\bibfnamefont {D.~M.}\ \bibnamefont {Paul}}, \bibinfo {author}
  {\bibfnamefont {G.}~\bibnamefont {Balakrishnan}}, \bibinfo {author}
  {\bibfnamefont {E.}~\bibnamefont {Pomjakushina}}, \bibinfo {author}
  {\bibfnamefont {M.}~\bibnamefont {Shi}}, \bibinfo {author} {\bibfnamefont
  {M.}~\bibnamefont {Medarde}}, \bibinfo {author} {\bibfnamefont {A.~D.}\
  \bibnamefont {Hillier}}, \bibinfo {author} {\bibfnamefont {H.~Q.}\
  \bibnamefont {Yuan}}, \bibinfo {author} {\bibfnamefont {J.}~\bibnamefont
  {Quintanilla}}, \bibinfo {author} {\bibfnamefont {J.}~\bibnamefont {Mesot}},\
  and\ \bibinfo {author} {\bibfnamefont {T.}~\bibnamefont {Shiroka}},\
  }\bibfield  {title} {\bibinfo {title} {Time-reversal symmetry breaking in
  {Re}-based superconductors},\ }\href
  {https://doi.org/10.1103/PhysRevLett.121.257002} {\bibfield  {journal}
  {\bibinfo  {journal} {Phys. Rev. Lett.}\ }\textbf {\bibinfo {volume} {121}},\
  \bibinfo {pages} {257002} (\bibinfo {year} {2018}{\natexlab{b}})}\BibitemShut
  {NoStop}%
\bibitem [{\citenamefont {Singh}\ \emph {et~al.}(2017)\citenamefont {Singh},
  \citenamefont {Barker}, \citenamefont {Thamizhavel}, \citenamefont {Paul},
  \citenamefont {Hillier},\ and\ \citenamefont {Singh}}]{Singh2017}%
  \BibitemOpen
  \bibfield  {author} {\bibinfo {author} {\bibfnamefont {D.}~\bibnamefont
  {Singh}}, \bibinfo {author} {\bibfnamefont {J.~A.~T.}\ \bibnamefont
  {Barker}}, \bibinfo {author} {\bibfnamefont {A.}~\bibnamefont {Thamizhavel}},
  \bibinfo {author} {\bibfnamefont {D.~M.}\ \bibnamefont {Paul}}, \bibinfo
  {author} {\bibfnamefont {A.~D.}\ \bibnamefont {Hillier}},\ and\ \bibinfo
  {author} {\bibfnamefont {R.~P.}\ \bibnamefont {Singh}},\ }\bibfield  {title}
  {\bibinfo {title} {Time-reversal symmetry breaking in the noncentrosymmetric
  superconductor {Re}$_{6}${Hf}: Further evidence for unconventional behavior
  in the $\alpha$-{Mn} family of materials},\ }\href
  {https://doi.org/10.1103/PhysRevB.96.180501} {\bibfield  {journal} {\bibinfo
  {journal} {Phys. Rev. B}\ }\textbf {\bibinfo {volume} {96}},\ \bibinfo
  {pages} {180501(R)} (\bibinfo {year} {2017})}\BibitemShut {NoStop}%
\bibitem [{\citenamefont {Mackenzie}\ and\ \citenamefont
  {Maeno}(2003)}]{Mackenzie2003}%
  \BibitemOpen
  \bibfield  {author} {\bibinfo {author} {\bibfnamefont {A.~P.}\ \bibnamefont
  {Mackenzie}}\ and\ \bibinfo {author} {\bibfnamefont {Y.}~\bibnamefont
  {Maeno}},\ }\bibfield  {title} {\bibinfo {title} {The superconductivity of
  {Sr}$_{2}${Ru}{O}$_{4}$ and the physics of spin-triplet pairing},\ }\href
  {https://doi.org/10.1103/RevModPhys.75.657} {\bibfield  {journal} {\bibinfo
  {journal} {Rev. Mod. Phys.}\ }\textbf {\bibinfo {volume} {75}},\ \bibinfo
  {pages} {657} (\bibinfo {year} {2003})},\ \bibinfo {note} {and reference
  therein}\BibitemShut {NoStop}%
\bibitem [{\citenamefont {Joynt}\ and\ \citenamefont
  {Taillefer}(2002)}]{Rober2002}%
  \BibitemOpen
  \bibfield  {author} {\bibinfo {author} {\bibfnamefont {R.}~\bibnamefont
  {Joynt}}\ and\ \bibinfo {author} {\bibfnamefont {L.}~\bibnamefont
  {Taillefer}},\ }\bibfield  {title} {\bibinfo {title} {The superconducting
  phases of {U}{Pt}$_3$},\ }\href {https://doi.org/10.1103/RevModPhys.74.235}
  {\bibfield  {journal} {\bibinfo  {journal} {Rev. Mod. Phys.}\ }\textbf
  {\bibinfo {volume} {74}},\ \bibinfo {pages} {235} (\bibinfo {year} {2002})},\
  \bibinfo {note} {and reference therein}\BibitemShut {NoStop}%
\bibitem [{\citenamefont {Bauer}\ and\ \citenamefont
  {Sigrist}(2012)}]{Bauer2012}%
  \BibitemOpen
  \bibinfo {editor} {\bibfnamefont {E.}~\bibnamefont {Bauer}}\ and\ \bibinfo
  {editor} {\bibfnamefont {M.}~\bibnamefont {Sigrist}},\ eds.,\ \href@noop {}
  {\emph {\bibinfo {title} {Non-Centrosymmetric Superconductors}}},\ Vol.\
  \bibinfo {volume} {847}\ (\bibinfo  {publisher} {Springer Verlag},\ \bibinfo
  {address} {Berlin},\ \bibinfo {year} {2012})\BibitemShut {NoStop}%
\bibitem [{\citenamefont {Sungkit}(2014)}]{Sungkit2014}%
  \BibitemOpen
  \bibfield  {author} {\bibinfo {author} {\bibfnamefont {Y.}~\bibnamefont
  {Sungkit}},\ }\bibfield  {title} {\bibinfo {title} {Noncentrosymmetric
  superconductors},\ }\href
  {https://doi.org/10.1146/annurev-conmatphys-031113-133912} {\bibfield
  {journal} {\bibinfo  {journal} {Annu. Rev. Condens. Matter Phys.}\ }\textbf
  {\bibinfo {volume} {5}},\ \bibinfo {pages} {15} (\bibinfo {year}
  {2014})}\BibitemShut {NoStop}%
\bibitem [{\citenamefont {Smidman}\ \emph {et~al.}(2017)\citenamefont
  {Smidman}, \citenamefont {Salamon}, \citenamefont {Yuan},\ and\ \citenamefont
  {Agterberg}}]{smidman2017}%
  \BibitemOpen
  \bibfield  {author} {\bibinfo {author} {\bibfnamefont {M.}~\bibnamefont
  {Smidman}}, \bibinfo {author} {\bibfnamefont {M.~B.}\ \bibnamefont
  {Salamon}}, \bibinfo {author} {\bibfnamefont {H.~Q.}\ \bibnamefont {Yuan}},\
  and\ \bibinfo {author} {\bibfnamefont {D.~F.}\ \bibnamefont {Agterberg}},\
  }\bibfield  {title} {\bibinfo {title} {Superconductivity and spin--orbit
  coupling in non-centrosymmetric materials: {A} review},\ }\href
  {http://stacks.iop.org/0034-4885/80/i=3/a=036501} {\bibfield  {journal}
  {\bibinfo  {journal} {Rep. Prog. Phys.}\ }\textbf {\bibinfo {volume} {80}},\
  \bibinfo {pages} {036501} (\bibinfo {year} {2017})}\BibitemShut {NoStop}%
\bibitem [{\citenamefont {Yuan}\ \emph {et~al.}(2006)\citenamefont {Yuan},
  \citenamefont {Agterberg}, \citenamefont {Hayashi}, \citenamefont {Badica},
  \citenamefont {Vandervelde}, \citenamefont {Togano}, \citenamefont
  {Sigrist},\ and\ \citenamefont {Salamon}}]{yuan2006}%
  \BibitemOpen
  \bibfield  {author} {\bibinfo {author} {\bibfnamefont {H.~Q.}\ \bibnamefont
  {Yuan}}, \bibinfo {author} {\bibfnamefont {D.~F.}\ \bibnamefont {Agterberg}},
  \bibinfo {author} {\bibfnamefont {N.}~\bibnamefont {Hayashi}}, \bibinfo
  {author} {\bibfnamefont {P.}~\bibnamefont {Badica}}, \bibinfo {author}
  {\bibfnamefont {D.}~\bibnamefont {Vandervelde}}, \bibinfo {author}
  {\bibfnamefont {K.}~\bibnamefont {Togano}}, \bibinfo {author} {\bibfnamefont
  {M.}~\bibnamefont {Sigrist}},\ and\ \bibinfo {author} {\bibfnamefont {M.~B.}\
  \bibnamefont {Salamon}},\ }\bibfield  {title} {\bibinfo {title} {S-wave
  spin-triplet order in superconductors without inversion symmetry:
  {Li}$_{2}${Pd}$_{3}${B} and {Li}$_{2}${Pt}$_{3}${B}},\ }\href
  {https://doi.org/10.1103/PhysRevLett.97.017006} {\bibfield  {journal}
  {\bibinfo  {journal} {Phys. Rev. Lett.}\ }\textbf {\bibinfo {volume} {97}},\
  \bibinfo {pages} {017006} (\bibinfo {year} {2006})}\BibitemShut {NoStop}%
\bibitem [{\citenamefont {Nishiyama}\ \emph {et~al.}(2007)\citenamefont
  {Nishiyama}, \citenamefont {Inada},\ and\ \citenamefont
  {Zheng}}]{nishiyama2007}%
  \BibitemOpen
  \bibfield  {author} {\bibinfo {author} {\bibfnamefont {M.}~\bibnamefont
  {Nishiyama}}, \bibinfo {author} {\bibfnamefont {Y.}~\bibnamefont {Inada}},\
  and\ \bibinfo {author} {\bibfnamefont {G.-q.}\ \bibnamefont {Zheng}},\
  }\bibfield  {title} {\bibinfo {title} {Spin triplet superconducting state due
  to broken inversion symmetry in {Li}$_{2}${Pt}$_{3}${B}},\ }\href
  {https://doi.org/10.1103/PhysRevLett.98.047002} {\bibfield  {journal}
  {\bibinfo  {journal} {Phys. Rev. Lett.}\ }\textbf {\bibinfo {volume} {98}},\
  \bibinfo {pages} {047002} (\bibinfo {year} {2007})}\BibitemShut {NoStop}%
\bibitem [{\citenamefont {Bonalde}\ \emph {et~al.}(2005)\citenamefont
  {Bonalde}, \citenamefont {Br{\"a}mer-Escamilla},\ and\ \citenamefont
  {Bauer}}]{bonalde2005CePt3Si}%
  \BibitemOpen
  \bibfield  {author} {\bibinfo {author} {\bibfnamefont {I.}~\bibnamefont
  {Bonalde}}, \bibinfo {author} {\bibfnamefont {W.}~\bibnamefont
  {Br{\"a}mer-Escamilla}},\ and\ \bibinfo {author} {\bibfnamefont
  {E.}~\bibnamefont {Bauer}},\ }\bibfield  {title} {\bibinfo {title} {Evidence
  for line nodes in the superconducting energy gap of noncentrosymmetric
  {Ce}{Pt}$_{3}${Si} from magnetic penetration depth measurements},\ }\href
  {https://doi.org/10.1103/PhysRevLett.94.207002} {\bibfield  {journal}
  {\bibinfo  {journal} {Phys. Rev. Lett.}\ }\textbf {\bibinfo {volume} {94}},\
  \bibinfo {pages} {207002} (\bibinfo {year} {2005})}\BibitemShut {NoStop}%
\bibitem [{\citenamefont {Pang}\ \emph {et~al.}(2015)\citenamefont {Pang},
  \citenamefont {Smidman}, \citenamefont {Jiang}, \citenamefont {Bao},
  \citenamefont {Weng}, \citenamefont {Wang}, \citenamefont {Jiao},
  \citenamefont {Zhang}, \citenamefont {Cao},\ and\ \citenamefont
  {Yuan}}]{K2Cr3As3Pen}%
  \BibitemOpen
  \bibfield  {author} {\bibinfo {author} {\bibfnamefont {G.~M.}\ \bibnamefont
  {Pang}}, \bibinfo {author} {\bibfnamefont {M.}~\bibnamefont {Smidman}},
  \bibinfo {author} {\bibfnamefont {W.~B.}\ \bibnamefont {Jiang}}, \bibinfo
  {author} {\bibfnamefont {J.~K.}\ \bibnamefont {Bao}}, \bibinfo {author}
  {\bibfnamefont {Z.~F.}\ \bibnamefont {Weng}}, \bibinfo {author}
  {\bibfnamefont {Y.~F.}\ \bibnamefont {Wang}}, \bibinfo {author}
  {\bibfnamefont {L.}~\bibnamefont {Jiao}}, \bibinfo {author} {\bibfnamefont
  {J.~L.}\ \bibnamefont {Zhang}}, \bibinfo {author} {\bibfnamefont {G.~H.}\
  \bibnamefont {Cao}},\ and\ \bibinfo {author} {\bibfnamefont {H.~Q.}\
  \bibnamefont {Yuan}},\ }\bibfield  {title} {\bibinfo {title} {Evidence for
  nodal superconductivity in quasi-one-dimensional
  {K}$_{2}${Cr}$_{3}${As}$_{3}$},\ }\href
  {https://doi.org/10.1103/PhysRevB.91.220502} {\bibfield  {journal} {\bibinfo
  {journal} {Phys. Rev. B}\ }\textbf {\bibinfo {volume} {91}},\ \bibinfo
  {pages} {220502(R)} (\bibinfo {year} {2015})}\BibitemShut {NoStop}%
\bibitem [{\citenamefont {Adroja}\ \emph {et~al.}(2015)\citenamefont {Adroja},
  \citenamefont {Bhattacharyya}, \citenamefont {Telling}, \citenamefont {Feng},
  \citenamefont {Smidman}, \citenamefont {Pan}, \citenamefont {Zhao},
  \citenamefont {Hillier}, \citenamefont {Pratt},\ and\ \citenamefont
  {Strydom}}]{K2Cr3As3MuSR}%
  \BibitemOpen
  \bibfield  {author} {\bibinfo {author} {\bibfnamefont {D.~T.}\ \bibnamefont
  {Adroja}}, \bibinfo {author} {\bibfnamefont {A.}~\bibnamefont
  {Bhattacharyya}}, \bibinfo {author} {\bibfnamefont {M.}~\bibnamefont
  {Telling}}, \bibinfo {author} {\bibfnamefont {Y.}~\bibnamefont {Feng}},
  \bibinfo {author} {\bibfnamefont {M.}~\bibnamefont {Smidman}}, \bibinfo
  {author} {\bibfnamefont {B.}~\bibnamefont {Pan}}, \bibinfo {author}
  {\bibfnamefont {J.}~\bibnamefont {Zhao}}, \bibinfo {author} {\bibfnamefont
  {A.~D.}\ \bibnamefont {Hillier}}, \bibinfo {author} {\bibfnamefont {F.~L.}\
  \bibnamefont {Pratt}},\ and\ \bibinfo {author} {\bibfnamefont {A.~M.}\
  \bibnamefont {Strydom}},\ }\bibfield  {title} {\bibinfo {title}
  {Superconducting ground state of quasi-one-dimensional
  {K}$_{2}${Cr}$_{3}${As}$_{3}$ investigated using $\ensuremath{\mu}\text{SR}$
  measurements},\ }\href {https://doi.org/10.1103/PhysRevB.92.134505}
  {\bibfield  {journal} {\bibinfo  {journal} {Phys. Rev. B}\ }\textbf {\bibinfo
  {volume} {92}},\ \bibinfo {pages} {134505} (\bibinfo {year}
  {2015})}\BibitemShut {NoStop}%
\bibitem [{\citenamefont {Bauer}\ \emph {et~al.}(2004)\citenamefont {Bauer},
  \citenamefont {Hilscher}, \citenamefont {Michor}, \citenamefont {Paul},
  \citenamefont {Scheidt}, \citenamefont {Gribanov}, \citenamefont {Seropegin},
  \citenamefont {No{\"e}l}, \citenamefont {Sigrist},\ and\ \citenamefont
  {Rogl}}]{bauer2004}%
  \BibitemOpen
  \bibfield  {author} {\bibinfo {author} {\bibfnamefont {E.}~\bibnamefont
  {Bauer}}, \bibinfo {author} {\bibfnamefont {G.}~\bibnamefont {Hilscher}},
  \bibinfo {author} {\bibfnamefont {H.}~\bibnamefont {Michor}}, \bibinfo
  {author} {\bibfnamefont {C.}~\bibnamefont {Paul}}, \bibinfo {author}
  {\bibfnamefont {E.~W.}\ \bibnamefont {Scheidt}}, \bibinfo {author}
  {\bibfnamefont {A.}~\bibnamefont {Gribanov}}, \bibinfo {author}
  {\bibfnamefont {Y.}~\bibnamefont {Seropegin}}, \bibinfo {author}
  {\bibfnamefont {H.}~\bibnamefont {No{\"e}l}}, \bibinfo {author}
  {\bibfnamefont {M.}~\bibnamefont {Sigrist}},\ and\ \bibinfo {author}
  {\bibfnamefont {P.}~\bibnamefont {Rogl}},\ }\bibfield  {title} {\bibinfo
  {title} {Heavy fermion superconductivity and magnetic order in
  noncentrosymmetric {Ce}{Pt}$_{3}${Si}},\ }\href
  {https://doi.org/10.1103/PhysRevLett.92.027003} {\bibfield  {journal}
  {\bibinfo  {journal} {Phys. Rev. Lett.}\ }\textbf {\bibinfo {volume} {92}},\
  \bibinfo {pages} {027003} (\bibinfo {year} {2004})}\BibitemShut {NoStop}%
\bibitem [{\citenamefont {Carnicom}\ \emph {et~al.}(2018)\citenamefont
  {Carnicom}, \citenamefont {Xie}, \citenamefont {Klimczuk}, \citenamefont
  {Lin}, \citenamefont {G{\'o}rnicka}, \citenamefont {Sobczak}, \citenamefont
  {Ong},\ and\ \citenamefont {Cava}}]{Carnicom2018}%
  \BibitemOpen
  \bibfield  {author} {\bibinfo {author} {\bibfnamefont {E.~M.}\ \bibnamefont
  {Carnicom}}, \bibinfo {author} {\bibfnamefont {W.~w.}\ \bibnamefont {Xie}},
  \bibinfo {author} {\bibfnamefont {T.}~\bibnamefont {Klimczuk}}, \bibinfo
  {author} {\bibfnamefont {J.~J.}\ \bibnamefont {Lin}}, \bibinfo {author}
  {\bibfnamefont {K.}~\bibnamefont {G{\'o}rnicka}}, \bibinfo {author}
  {\bibfnamefont {Z.}~\bibnamefont {Sobczak}}, \bibinfo {author} {\bibfnamefont
  {N.~P.}\ \bibnamefont {Ong}},\ and\ \bibinfo {author} {\bibfnamefont {R.~J.}\
  \bibnamefont {Cava}},\ }\bibfield  {title} {\bibinfo {title} {The chiral
  non-centrosymmetric superconductors {Ta}{Rh}$_{2}${B}$_{2}$ and nbrh
  {Nb}{Rh}$_{2}${B}$_{2}$},\ }\href {https://doi.org/10.1126/sciadv.aar7969}
  {\bibfield  {journal} {\bibinfo  {journal} {Sci. Adv.}\ }\textbf {\bibinfo
  {volume} {4}},\ \bibinfo {pages} {eaar7969} (\bibinfo {year}
  {2018})}\BibitemShut {NoStop}%
\bibitem [{\citenamefont {Kimura}\ \emph {et~al.}(2007)\citenamefont {Kimura},
  \citenamefont {Ito}, \citenamefont {Aoki}, \citenamefont {Uji},\ and\
  \citenamefont {Terashima}}]{kimura2007}%
  \BibitemOpen
  \bibfield  {author} {\bibinfo {author} {\bibfnamefont {N.}~\bibnamefont
  {Kimura}}, \bibinfo {author} {\bibfnamefont {K.}~\bibnamefont {Ito}},
  \bibinfo {author} {\bibfnamefont {H.}~\bibnamefont {Aoki}}, \bibinfo {author}
  {\bibfnamefont {S.}~\bibnamefont {Uji}},\ and\ \bibinfo {author}
  {\bibfnamefont {T.}~\bibnamefont {Terashima}},\ }\bibfield  {title} {\bibinfo
  {title} {Extremely high upper critical magnetic field of the
  noncentrosymmetric heavy fermion superconductor {Ce}{Rh}{Si}$_{3}$},\ }\href
  {https://doi.org/10.1103/PhysRevLett.98.197001} {\bibfield  {journal}
  {\bibinfo  {journal} {Phys. Rev. Lett.}\ }\textbf {\bibinfo {volume} {98}},\
  \bibinfo {pages} {197001} (\bibinfo {year} {2007})}\BibitemShut {NoStop}%
\bibitem [{\citenamefont {Chen}\ \emph {et~al.}(2011)\citenamefont {Chen},
  \citenamefont {Salamon}, \citenamefont {Akutagawa}, \citenamefont {Akimitsu},
  \citenamefont {Singleton}, \citenamefont {Zhang}, \citenamefont {Jiao},\ and\
  \citenamefont {Yuan}}]{Chen2011}%
  \BibitemOpen
  \bibfield  {author} {\bibinfo {author} {\bibfnamefont {J.}~\bibnamefont
  {Chen}}, \bibinfo {author} {\bibfnamefont {M.~B.}\ \bibnamefont {Salamon}},
  \bibinfo {author} {\bibfnamefont {S.}~\bibnamefont {Akutagawa}}, \bibinfo
  {author} {\bibfnamefont {J.}~\bibnamefont {Akimitsu}}, \bibinfo {author}
  {\bibfnamefont {J.}~\bibnamefont {Singleton}}, \bibinfo {author}
  {\bibfnamefont {J.~L.}\ \bibnamefont {Zhang}}, \bibinfo {author}
  {\bibfnamefont {L.}~\bibnamefont {Jiao}},\ and\ \bibinfo {author}
  {\bibfnamefont {H.~Q.}\ \bibnamefont {Yuan}},\ }\bibfield  {title} {\bibinfo
  {title} {Evidence of nodal gap structure in the noncentrosymmetric
  superconductor {Y}$_2${C}$_3$},\ }\href
  {https://doi.org/10.1103/PhysRevB.83.144529} {\bibfield  {journal} {\bibinfo
  {journal} {Phys. Rev. B}\ }\textbf {\bibinfo {volume} {83}},\ \bibinfo
  {pages} {144529} (\bibinfo {year} {2011})}\BibitemShut {NoStop}%
\bibitem [{\citenamefont {Kim}\ \emph {et~al.}(2018)\citenamefont {Kim},
  \citenamefont {Wang}, \citenamefont {Nakajima}, \citenamefont {Hu},
  \citenamefont {Ziemak}, \citenamefont {Syers}, \citenamefont {Wang},
  \citenamefont {Hodovanets}, \citenamefont {Denlinger}, \citenamefont
  {Brydon}, \citenamefont {Agterberg}, \citenamefont {Tanatar}, \citenamefont
  {Prozorov},\ and\ \citenamefont {Paglione}}]{Kim2018}%
  \BibitemOpen
  \bibfield  {author} {\bibinfo {author} {\bibfnamefont {H.}~\bibnamefont
  {Kim}}, \bibinfo {author} {\bibfnamefont {K.}~\bibnamefont {Wang}}, \bibinfo
  {author} {\bibfnamefont {Y.}~\bibnamefont {Nakajima}}, \bibinfo {author}
  {\bibfnamefont {R.}~\bibnamefont {Hu}}, \bibinfo {author} {\bibfnamefont
  {S.}~\bibnamefont {Ziemak}}, \bibinfo {author} {\bibfnamefont
  {P.}~\bibnamefont {Syers}}, \bibinfo {author} {\bibfnamefont
  {L.}~\bibnamefont {Wang}}, \bibinfo {author} {\bibfnamefont {H.}~\bibnamefont
  {Hodovanets}}, \bibinfo {author} {\bibfnamefont {J.~D.}\ \bibnamefont
  {Denlinger}}, \bibinfo {author} {\bibfnamefont {P.~M.~R.}\ \bibnamefont
  {Brydon}}, \bibinfo {author} {\bibfnamefont {D.~F.}\ \bibnamefont
  {Agterberg}}, \bibinfo {author} {\bibfnamefont {M.~A.}\ \bibnamefont
  {Tanatar}}, \bibinfo {author} {\bibfnamefont {R.}~\bibnamefont {Prozorov}},\
  and\ \bibinfo {author} {\bibfnamefont {J.}~\bibnamefont {Paglione}},\
  }\bibfield  {title} {\bibinfo {title} {Beyond triplet: Unconventional
  superconductivity in a spin-$3/2$ topological semimetal},\ }\href
  {https://doi.org/10.1126/sciadv.aao4513} {\bibfield  {journal} {\bibinfo
  {journal} {Sci. Adv.}\ }\textbf {\bibinfo {volume} {4}},\ \bibinfo {pages}
  {eaao4513} (\bibinfo {year} {2018})}\BibitemShut {NoStop}%
\bibitem [{\citenamefont {Sun}\ \emph {et~al.}(2015)\citenamefont {Sun},
  \citenamefont {Enayat}, \citenamefont {Maldonado}, \citenamefont {Lithgow},
  \citenamefont {Yelland}, \citenamefont {Peets}, \citenamefont {Yaresko},
  \citenamefont {Schnyder},\ and\ \citenamefont {Wahl}}]{Sun2015}%
  \BibitemOpen
  \bibfield  {author} {\bibinfo {author} {\bibfnamefont {Z.~X.}\ \bibnamefont
  {Sun}}, \bibinfo {author} {\bibfnamefont {M.}~\bibnamefont {Enayat}},
  \bibinfo {author} {\bibfnamefont {A.}~\bibnamefont {Maldonado}}, \bibinfo
  {author} {\bibfnamefont {C.}~\bibnamefont {Lithgow}}, \bibinfo {author}
  {\bibfnamefont {E.}~\bibnamefont {Yelland}}, \bibinfo {author} {\bibfnamefont
  {D.~C.}\ \bibnamefont {Peets}}, \bibinfo {author} {\bibfnamefont
  {A.}~\bibnamefont {Yaresko}}, \bibinfo {author} {\bibfnamefont {A.~P.}\
  \bibnamefont {Schnyder}},\ and\ \bibinfo {author} {\bibfnamefont
  {P.}~\bibnamefont {Wahl}},\ }\bibfield  {title} {\bibinfo {title} {Dirac
  surface states and nature of superconductivity in noncentrosymmetric
  {Bi}{Pd}},\ }\href {https://doi.org/10.1038/ncomms7633} {\bibfield  {journal}
  {\bibinfo  {journal} {Nat. Commun.}\ }\textbf {\bibinfo {volume} {6}},\
  \bibinfo {pages} {6633} (\bibinfo {year} {2015})}\BibitemShut {NoStop}%
\bibitem [{\citenamefont {Ali}\ \emph {et~al.}(2014)\citenamefont {Ali},
  \citenamefont {Gibson}, \citenamefont {Klimczuk},\ and\ \citenamefont
  {Cava}}]{Ali2014}%
  \BibitemOpen
  \bibfield  {author} {\bibinfo {author} {\bibfnamefont {M.~N.}\ \bibnamefont
  {Ali}}, \bibinfo {author} {\bibfnamefont {Q.~D.}\ \bibnamefont {Gibson}},
  \bibinfo {author} {\bibfnamefont {T.}~\bibnamefont {Klimczuk}},\ and\
  \bibinfo {author} {\bibfnamefont {R.~J.}\ \bibnamefont {Cava}},\ }\bibfield
  {title} {\bibinfo {title} {Noncentrosymmetric superconductor with a bulk
  three-dimensional {D}irac cone gapped by strong spin-orbit coupling},\ }\href
  {https://doi.org/10.1103/PhysRevB.89.020505} {\bibfield  {journal} {\bibinfo
  {journal} {Phys. Rev. B}\ }\textbf {\bibinfo {volume} {89}},\ \bibinfo
  {pages} {020505(R)} (\bibinfo {year} {2014})}\BibitemShut {NoStop}%
\bibitem [{\citenamefont {Sato}\ and\ \citenamefont
  {Fujimoto}(2009)}]{Sato2009}%
  \BibitemOpen
  \bibfield  {author} {\bibinfo {author} {\bibfnamefont {M.}~\bibnamefont
  {Sato}}\ and\ \bibinfo {author} {\bibfnamefont {S.}~\bibnamefont
  {Fujimoto}},\ }\bibfield  {title} {\bibinfo {title} {Topological phases of
  noncentrosymmetric superconductors: {E}dge states, {M}ajorana fermions, and
  non-{A}belian statistics},\ }\href
  {https://doi.org/10.1103/PhysRevB.79.094504} {\bibfield  {journal} {\bibinfo
  {journal} {Phys. Rev. B}\ }\textbf {\bibinfo {volume} {79}},\ \bibinfo
  {pages} {094504} (\bibinfo {year} {2009})}\BibitemShut {NoStop}%
\bibitem [{\citenamefont {Tanaka}\ \emph {et~al.}(2010)\citenamefont {Tanaka},
  \citenamefont {Mizuno}, \citenamefont {Yokoyama}, \citenamefont {Yada},\ and\
  \citenamefont {Sato}}]{Tanaka2010}%
  \BibitemOpen
  \bibfield  {author} {\bibinfo {author} {\bibfnamefont {Y.}~\bibnamefont
  {Tanaka}}, \bibinfo {author} {\bibfnamefont {Y.}~\bibnamefont {Mizuno}},
  \bibinfo {author} {\bibfnamefont {T.}~\bibnamefont {Yokoyama}}, \bibinfo
  {author} {\bibfnamefont {K.}~\bibnamefont {Yada}},\ and\ \bibinfo {author}
  {\bibfnamefont {M.}~\bibnamefont {Sato}},\ }\bibfield  {title} {\bibinfo
  {title} {Anomalous {A}ndreev bound state in noncentrosymmetric
  superconductors},\ }\href {https://doi.org/10.1103/PhysRevLett.105.097002}
  {\bibfield  {journal} {\bibinfo  {journal} {Phys. Rev. Lett.}\ }\textbf
  {\bibinfo {volume} {105}},\ \bibinfo {pages} {097002} (\bibinfo {year}
  {2010})}\BibitemShut {NoStop}%
\bibitem [{\citenamefont {Bauer}\ \emph {et~al.}(2010)\citenamefont {Bauer},
  \citenamefont {Rogl}, \citenamefont {Chen}, \citenamefont {Khan},
  \citenamefont {Michor}, \citenamefont {Hilscher}, \citenamefont {Royanian},
  \citenamefont {Kumagai}, \citenamefont {Li}, \citenamefont {Li},
  \citenamefont {Podloucky},\ and\ \citenamefont {Rogl}}]{bauer2010}%
  \BibitemOpen
  \bibfield  {author} {\bibinfo {author} {\bibfnamefont {E.}~\bibnamefont
  {Bauer}}, \bibinfo {author} {\bibfnamefont {G.}~\bibnamefont {Rogl}},
  \bibinfo {author} {\bibfnamefont {X.-Q.}\ \bibnamefont {Chen}}, \bibinfo
  {author} {\bibfnamefont {R.~T.}\ \bibnamefont {Khan}}, \bibinfo {author}
  {\bibfnamefont {H.}~\bibnamefont {Michor}}, \bibinfo {author} {\bibfnamefont
  {G.}~\bibnamefont {Hilscher}}, \bibinfo {author} {\bibfnamefont
  {E.}~\bibnamefont {Royanian}}, \bibinfo {author} {\bibfnamefont
  {K.}~\bibnamefont {Kumagai}}, \bibinfo {author} {\bibfnamefont {D.~Z.}\
  \bibnamefont {Li}}, \bibinfo {author} {\bibfnamefont {Y.~Y.}\ \bibnamefont
  {Li}}, \bibinfo {author} {\bibfnamefont {R.}~\bibnamefont {Podloucky}},\ and\
  \bibinfo {author} {\bibfnamefont {P.}~\bibnamefont {Rogl}},\ }\bibfield
  {title} {\bibinfo {title} {Unconventional superconducting phase in the weakly
  correlated noncentrosymmetric {Mo}$_{3}${Al}$_{2}${C} compound},\ }\href
  {https://doi.org/10.1103/PhysRevB.82.064511} {\bibfield  {journal} {\bibinfo
  {journal} {Phys. Rev. B}\ }\textbf {\bibinfo {volume} {82}},\ \bibinfo
  {pages} {064511} (\bibinfo {year} {2010})}\BibitemShut {NoStop}%
\bibitem [{\citenamefont {Anand}\ \emph {et~al.}(2011)\citenamefont {Anand},
  \citenamefont {Hillier}, \citenamefont {Adroja}, \citenamefont {Strydom},
  \citenamefont {Michor}, \citenamefont {McEwen},\ and\ \citenamefont
  {Rainford}}]{Anand2011}%
  \BibitemOpen
  \bibfield  {author} {\bibinfo {author} {\bibfnamefont {V.~K.}\ \bibnamefont
  {Anand}}, \bibinfo {author} {\bibfnamefont {A.~D.}\ \bibnamefont {Hillier}},
  \bibinfo {author} {\bibfnamefont {D.~T.}\ \bibnamefont {Adroja}}, \bibinfo
  {author} {\bibfnamefont {A.~M.}\ \bibnamefont {Strydom}}, \bibinfo {author}
  {\bibfnamefont {H.}~\bibnamefont {Michor}}, \bibinfo {author} {\bibfnamefont
  {K.~A.}\ \bibnamefont {McEwen}},\ and\ \bibinfo {author} {\bibfnamefont
  {B.~D.}\ \bibnamefont {Rainford}},\ }\bibfield  {title} {\bibinfo {title}
  {Specific heat and $\mu$sr study on the noncentrosymmetric superconductor
  {La}{Rh}{Si}$_{3}$},\ }\href {https://doi.org/10.1103/PhysRevB.83.064522}
  {\bibfield  {journal} {\bibinfo  {journal} {Phys. Rev. B}\ }\textbf {\bibinfo
  {volume} {83}},\ \bibinfo {pages} {064522} (\bibinfo {year}
  {2011})}\BibitemShut {NoStop}%
\bibitem [{\citenamefont {Anand}\ \emph {et~al.}(2014)\citenamefont {Anand},
  \citenamefont {Britz}, \citenamefont {Bhattacharyya}, \citenamefont {Adroja},
  \citenamefont {Hillier}, \citenamefont {Strydom}, \citenamefont {Kockelmann},
  \citenamefont {Rainford},\ and\ \citenamefont {McEwen}}]{Anand2014}%
  \BibitemOpen
  \bibfield  {author} {\bibinfo {author} {\bibfnamefont {V.~K.}\ \bibnamefont
  {Anand}}, \bibinfo {author} {\bibfnamefont {D.}~\bibnamefont {Britz}},
  \bibinfo {author} {\bibfnamefont {A.}~\bibnamefont {Bhattacharyya}}, \bibinfo
  {author} {\bibfnamefont {D.~T.}\ \bibnamefont {Adroja}}, \bibinfo {author}
  {\bibfnamefont {A.~D.}\ \bibnamefont {Hillier}}, \bibinfo {author}
  {\bibfnamefont {A.~M.}\ \bibnamefont {Strydom}}, \bibinfo {author}
  {\bibfnamefont {W.}~\bibnamefont {Kockelmann}}, \bibinfo {author}
  {\bibfnamefont {B.~D.}\ \bibnamefont {Rainford}},\ and\ \bibinfo {author}
  {\bibfnamefont {K.~A.}\ \bibnamefont {McEwen}},\ }\bibfield  {title}
  {\bibinfo {title} {Physical properties of noncentrosymmetric superconductor
  {La}{Ir}{Si}$_3$: {A} $\mu${SR} study},\ }\href
  {https://doi.org/10.1103/PhysRevB.90.014513} {\bibfield  {journal} {\bibinfo
  {journal} {Phys. Rev. B}\ }\textbf {\bibinfo {volume} {90}},\ \bibinfo
  {pages} {014513} (\bibinfo {year} {2014})}\BibitemShut {NoStop}%
\bibitem [{\citenamefont {Smidman}\ \emph {et~al.}(2014)\citenamefont
  {Smidman}, \citenamefont {Hillier}, \citenamefont {Adroja}, \citenamefont
  {Lees}, \citenamefont {Anand}, \citenamefont {Singh}, \citenamefont {Smith},
  \citenamefont {Paul},\ and\ \citenamefont {Balakrishnan}}]{Smidman2014}%
  \BibitemOpen
  \bibfield  {author} {\bibinfo {author} {\bibfnamefont {M.}~\bibnamefont
  {Smidman}}, \bibinfo {author} {\bibfnamefont {A.~D.}\ \bibnamefont
  {Hillier}}, \bibinfo {author} {\bibfnamefont {D.~T.}\ \bibnamefont {Adroja}},
  \bibinfo {author} {\bibfnamefont {M.~R.}\ \bibnamefont {Lees}}, \bibinfo
  {author} {\bibfnamefont {V.~K.}\ \bibnamefont {Anand}}, \bibinfo {author}
  {\bibfnamefont {R.~P.}\ \bibnamefont {Singh}}, \bibinfo {author}
  {\bibfnamefont {R.~I.}\ \bibnamefont {Smith}}, \bibinfo {author}
  {\bibfnamefont {D.~M.}\ \bibnamefont {Paul}},\ and\ \bibinfo {author}
  {\bibfnamefont {G.}~\bibnamefont {Balakrishnan}},\ }\bibfield  {title}
  {\bibinfo {title} {Investigations of the superconducting states of
  noncentrosymmetric {La}{Pd}{Si}$_{3}$ and {La}{Pt}{Si}$_{3}$},\ }\href
  {https://doi.org/10.1103/PhysRevB.89.094509} {\bibfield  {journal} {\bibinfo
  {journal} {Phys. Rev. B}\ }\textbf {\bibinfo {volume} {89}},\ \bibinfo
  {pages} {094509} (\bibinfo {year} {2014})}\BibitemShut {NoStop}%
\bibitem [{\citenamefont {Aczel}\ \emph {et~al.}(2010)\citenamefont {Aczel},
  \citenamefont {Williams}, \citenamefont {Goko}, \citenamefont {Carlo},
  \citenamefont {Yu}, \citenamefont {Uemura}, \citenamefont {Klimczuk},
  \citenamefont {Thompson}, \citenamefont {Cava},\ and\ \citenamefont
  {Luke}}]{Acze2010}%
  \BibitemOpen
  \bibfield  {author} {\bibinfo {author} {\bibfnamefont {A.~A.}\ \bibnamefont
  {Aczel}}, \bibinfo {author} {\bibfnamefont {T.~J.}\ \bibnamefont {Williams}},
  \bibinfo {author} {\bibfnamefont {T.}~\bibnamefont {Goko}}, \bibinfo {author}
  {\bibfnamefont {J.~P.}\ \bibnamefont {Carlo}}, \bibinfo {author}
  {\bibfnamefont {W.}~\bibnamefont {Yu}}, \bibinfo {author} {\bibfnamefont
  {Y.~J.}\ \bibnamefont {Uemura}}, \bibinfo {author} {\bibfnamefont
  {T.}~\bibnamefont {Klimczuk}}, \bibinfo {author} {\bibfnamefont {J.~D.}\
  \bibnamefont {Thompson}}, \bibinfo {author} {\bibfnamefont {R.~J.}\
  \bibnamefont {Cava}},\ and\ \bibinfo {author} {\bibfnamefont {G.~M.}\
  \bibnamefont {Luke}},\ }\bibfield  {title} {\bibinfo {title} {Muon spin
  rotation/relaxation measurements of the noncentrosymmetric superconductor
  {Mg}$_{10}${Ir}$_{19}${B}$_{16}$},\ }\href
  {https://doi.org/10.1103/PhysRevB.82.024520} {\bibfield  {journal} {\bibinfo
  {journal} {Phys. Rev. B}\ }\textbf {\bibinfo {volume} {82}},\ \bibinfo
  {pages} {024520} (\bibinfo {year} {2010})}\BibitemShut {NoStop}%
\bibitem [{\citenamefont {Shang}\ \emph {et~al.}(2019)\citenamefont {Shang},
  \citenamefont {Philippe}, \citenamefont {Verezhak}, \citenamefont {Guguchia},
  \citenamefont {Zhao}, \citenamefont {Chang}, \citenamefont {Lee},
  \citenamefont {Gawryluk}, \citenamefont {Pomjakushina}, \citenamefont {Shi},
  \citenamefont {Medarde}, \citenamefont {Ott},\ and\ \citenamefont
  {Shiroka}}]{Shang2019}%
  \BibitemOpen
  \bibfield  {author} {\bibinfo {author} {\bibfnamefont {T.}~\bibnamefont
  {Shang}}, \bibinfo {author} {\bibfnamefont {J.}~\bibnamefont {Philippe}},
  \bibinfo {author} {\bibfnamefont {J.~A.~T.}\ \bibnamefont {Verezhak}},
  \bibinfo {author} {\bibfnamefont {Z.}~\bibnamefont {Guguchia}}, \bibinfo
  {author} {\bibfnamefont {J.~Z.}\ \bibnamefont {Zhao}}, \bibinfo {author}
  {\bibfnamefont {L.-J.}\ \bibnamefont {Chang}}, \bibinfo {author}
  {\bibfnamefont {M.~K.}\ \bibnamefont {Lee}}, \bibinfo {author} {\bibfnamefont
  {D.~J.}\ \bibnamefont {Gawryluk}}, \bibinfo {author} {\bibfnamefont
  {E.}~\bibnamefont {Pomjakushina}}, \bibinfo {author} {\bibfnamefont
  {M.}~\bibnamefont {Shi}}, \bibinfo {author} {\bibfnamefont {M.}~\bibnamefont
  {Medarde}}, \bibinfo {author} {\bibfnamefont {H.-R.}\ \bibnamefont {Ott}},\
  and\ \bibinfo {author} {\bibfnamefont {T.}~\bibnamefont {Shiroka}},\
  }\bibfield  {title} {\bibinfo {title} {Nodeless superconductivity and
  preserved time-reversal symmetry in the noncentrosymmetric {Mo}$_3${P}
  superconductor},\ }\href {https://doi.org/10.1103/PhysRevB.99.184513}
  {\bibfield  {journal} {\bibinfo  {journal} {Phys. Rev. B}\ }\textbf {\bibinfo
  {volume} {99}},\ \bibinfo {pages} {184513} (\bibinfo {year}
  {2019})}\BibitemShut {NoStop}%
\bibitem [{\citenamefont {Lee}\ \emph {et~al.}(1996)\citenamefont {Lee},
  \citenamefont {Zeng}, \citenamefont {Yao},\ and\ \citenamefont
  {Chen}}]{Lee1996}%
  \BibitemOpen
  \bibfield  {author} {\bibinfo {author} {\bibfnamefont {W.~H.}\ \bibnamefont
  {Lee}}, \bibinfo {author} {\bibfnamefont {H.~K.}\ \bibnamefont {Zeng}},
  \bibinfo {author} {\bibfnamefont {Y.~D.}\ \bibnamefont {Yao}},\ and\ \bibinfo
  {author} {\bibfnamefont {Y.~Y.}\ \bibnamefont {Chen}},\ }\bibfield  {title}
  {\bibinfo {title} {Superconductivity in the {Ni} based ternary carbide
  {LaNiC$_2$}},\ }\href
  {https://doi.org/https://doi.org/10.1016/0921-4534(96)00309-7} {\bibfield
  {journal} {\bibinfo  {journal} {Physica C}\ }\textbf {\bibinfo {volume}
  {266}},\ \bibinfo {pages} {138} (\bibinfo {year} {1996})}\BibitemShut
  {NoStop}%
\bibitem [{\citenamefont {Chen}\ \emph {et~al.}(2013)\citenamefont {Chen},
  \citenamefont {Jiao}, \citenamefont {Zhang}, \citenamefont {Chen},
  \citenamefont {Yang}, \citenamefont {Nicklas}, \citenamefont {Steglich},\
  and\ \citenamefont {Yuan}}]{chen2013}%
  \BibitemOpen
  \bibfield  {author} {\bibinfo {author} {\bibfnamefont {J.}~\bibnamefont
  {Chen}}, \bibinfo {author} {\bibfnamefont {L.}~\bibnamefont {Jiao}}, \bibinfo
  {author} {\bibfnamefont {J.~L.}\ \bibnamefont {Zhang}}, \bibinfo {author}
  {\bibfnamefont {Y.}~\bibnamefont {Chen}}, \bibinfo {author} {\bibfnamefont
  {L.}~\bibnamefont {Yang}}, \bibinfo {author} {\bibfnamefont {M.}~\bibnamefont
  {Nicklas}}, \bibinfo {author} {\bibfnamefont {F.}~\bibnamefont {Steglich}},\
  and\ \bibinfo {author} {\bibfnamefont {H.~Q.}\ \bibnamefont {Yuan}},\
  }\bibfield  {title} {\bibinfo {title} {Evidence for two-gap superconductivity
  in the non-centrosymmetric compound {La}{Ni}{C}$_{2}$},\ }\href
  {http://stacks.iop.org/1367-2630/15/i=5/a=053005} {\bibfield  {journal}
  {\bibinfo  {journal} {New J. Phys.}\ }\textbf {\bibinfo {volume} {15}},\
  \bibinfo {pages} {053005} (\bibinfo {year} {2013})}\BibitemShut {NoStop}%
\bibitem [{\citenamefont {Hirose}\ \emph {et~al.}(2012)\citenamefont {Hirose},
  \citenamefont {Kishino}, \citenamefont {Sakaguchi}, \citenamefont {Miura},
  \citenamefont {Honda}, \citenamefont {Takeuchi}, \citenamefont {Yamamoto},
  \citenamefont {Haga}, \citenamefont {Harima}, \citenamefont {Settai},\ and\
  \citenamefont {\={O}nuki}}]{Hirose2012}%
  \BibitemOpen
  \bibfield  {author} {\bibinfo {author} {\bibfnamefont {Y.}~\bibnamefont
  {Hirose}}, \bibinfo {author} {\bibfnamefont {T.}~\bibnamefont {Kishino}},
  \bibinfo {author} {\bibfnamefont {J.}~\bibnamefont {Sakaguchi}}, \bibinfo
  {author} {\bibfnamefont {Y.}~\bibnamefont {Miura}}, \bibinfo {author}
  {\bibfnamefont {F.}~\bibnamefont {Honda}}, \bibinfo {author} {\bibfnamefont
  {T.}~\bibnamefont {Takeuchi}}, \bibinfo {author} {\bibfnamefont
  {E.}~\bibnamefont {Yamamoto}}, \bibinfo {author} {\bibfnamefont
  {Y.}~\bibnamefont {Haga}}, \bibinfo {author} {\bibfnamefont {H.}~\bibnamefont
  {Harima}}, \bibinfo {author} {\bibfnamefont {R.}~\bibnamefont {Settai}},\
  and\ \bibinfo {author} {\bibfnamefont {Y.}~\bibnamefont {\={O}nuki}},\
  }\bibfield  {title} {\bibinfo {title} {{Fermi Surface and Superconducting
  Properties of Non-centrosymmetric LaNiC$_2$}},\ }\href
  {https://doi.org/10.1143/JPSJ.81.113703} {\bibfield  {journal} {\bibinfo
  {journal} {J. Phys. Soc. Jpn.}\ }\textbf {\bibinfo {volume} {81}},\ \bibinfo
  {pages} {113703} (\bibinfo {year} {2012})}\BibitemShut {NoStop}%
\bibitem [{\citenamefont {Bonalde}\ \emph {et~al.}(2011)\citenamefont
  {Bonalde}, \citenamefont {Ribeiro}, \citenamefont {Syu}, \citenamefont
  {Sung},\ and\ \citenamefont {Lee}}]{Bonalde2011}%
  \BibitemOpen
  \bibfield  {author} {\bibinfo {author} {\bibfnamefont {I.}~\bibnamefont
  {Bonalde}}, \bibinfo {author} {\bibfnamefont {R.~L.}\ \bibnamefont
  {Ribeiro}}, \bibinfo {author} {\bibfnamefont {K.~J.}\ \bibnamefont {Syu}},
  \bibinfo {author} {\bibfnamefont {H.~H.}\ \bibnamefont {Sung}},\ and\
  \bibinfo {author} {\bibfnamefont {W.~H.}\ \bibnamefont {Lee}},\ }\bibfield
  {title} {\bibinfo {title} {Nodal gap structure in the noncentrosymmetric
  superconductor {LaNiC}$_2$ from magnetic-penetration-depth measurements},\
  }\href {https://doi.org/10.1088/1367-2630/13/12/123022} {\bibfield  {journal}
  {\bibinfo  {journal} {New J. Phys.}\ }\textbf {\bibinfo {volume} {13}},\
  \bibinfo {pages} {123022} (\bibinfo {year} {2011})}\BibitemShut {NoStop}%
\bibitem [{\citenamefont {Landaeta}\ \emph {et~al.}(2017)\citenamefont
  {Landaeta}, \citenamefont {Subero}, \citenamefont {Machado}, \citenamefont
  {Honda},\ and\ \citenamefont {Bonalde}}]{Landaeta2017}%
  \BibitemOpen
  \bibfield  {author} {\bibinfo {author} {\bibfnamefont {J.~F.}\ \bibnamefont
  {Landaeta}}, \bibinfo {author} {\bibfnamefont {D.}~\bibnamefont {Subero}},
  \bibinfo {author} {\bibfnamefont {P.}~\bibnamefont {Machado}}, \bibinfo
  {author} {\bibfnamefont {F.}~\bibnamefont {Honda}},\ and\ \bibinfo {author}
  {\bibfnamefont {I.}~\bibnamefont {Bonalde}},\ }\bibfield  {title} {\bibinfo
  {title} {Unconventional superconductivity and an ambient-pressure magnetic
  quantum critical point in single-crystal {LaNiC}$_2$},\ }\href
  {https://doi.org/10.1103/PhysRevB.96.174515} {\bibfield  {journal} {\bibinfo
  {journal} {Phys. Rev. B}\ }\textbf {\bibinfo {volume} {96}},\ \bibinfo
  {pages} {174515} (\bibinfo {year} {2017})}\BibitemShut {NoStop}%
\bibitem [{Note1()}]{Note1}%
  \BibitemOpen
  \bibinfo {note} {In an early report of the specific heat of LaNiC$_2$, a
  $T^2$-dependence of C/T was observed (indicating nodal SC)~\cite {Lee1996},
  but more recently exponential behavior of $C/T$ was reported, consistent with
  a fully-gapped superconducting state~\cite {chen2013,Hirose2012}. Similar
  inconsistencies are also found from magnetic penetration depth $\lambda (T)$
  measurements, where both a $T^2$- and an exponential temperature dependence
  have been reported~\cite {chen2013,Bonalde2011,Landaeta2017}, consistent with
  the presence of point nodes and fully gapped behavior,
  respectively.}\BibitemShut {Stop}%
\bibitem [{\citenamefont {Quintanilla}\ \emph {et~al.}(2010)\citenamefont
  {Quintanilla}, \citenamefont {Hillier}, \citenamefont {Annett},\ and\
  \citenamefont {Cywinski}}]{Quintanilla2010}%
  \BibitemOpen
  \bibfield  {author} {\bibinfo {author} {\bibfnamefont {J.}~\bibnamefont
  {Quintanilla}}, \bibinfo {author} {\bibfnamefont {A.~D.}\ \bibnamefont
  {Hillier}}, \bibinfo {author} {\bibfnamefont {J.~F.}\ \bibnamefont
  {Annett}},\ and\ \bibinfo {author} {\bibfnamefont {R.}~\bibnamefont
  {Cywinski}},\ }\bibfield  {title} {\bibinfo {title} {Relativistic analysis of
  the pairing symmetry of the noncentrosymmetric superconductor
  {La}{Ni}{C}$_2$},\ }\href {https://doi.org/10.1103/PhysRevB.82.174511}
  {\bibfield  {journal} {\bibinfo  {journal} {Phys. Rev. B}\ }\textbf {\bibinfo
  {volume} {82}},\ \bibinfo {pages} {174511} (\bibinfo {year}
  {2010})}\BibitemShut {NoStop}%
\bibitem [{\citenamefont {Weng}\ \emph {et~al.}(2016)\citenamefont {Weng},
  \citenamefont {Zhang}, \citenamefont {Smidman}, \citenamefont {Shang},
  \citenamefont {Quintanilla}, \citenamefont {Annett}, \citenamefont {Nicklas},
  \citenamefont {Pang}, \citenamefont {Jiao}, \citenamefont {Jiang},
  \citenamefont {Chen}, \citenamefont {Steglich},\ and\ \citenamefont
  {Yuan}}]{Weng2016}%
  \BibitemOpen
  \bibfield  {author} {\bibinfo {author} {\bibfnamefont {Z.~F.}\ \bibnamefont
  {Weng}}, \bibinfo {author} {\bibfnamefont {J.~L.}\ \bibnamefont {Zhang}},
  \bibinfo {author} {\bibfnamefont {M.}~\bibnamefont {Smidman}}, \bibinfo
  {author} {\bibfnamefont {T.}~\bibnamefont {Shang}}, \bibinfo {author}
  {\bibfnamefont {J.}~\bibnamefont {Quintanilla}}, \bibinfo {author}
  {\bibfnamefont {J.~F.}\ \bibnamefont {Annett}}, \bibinfo {author}
  {\bibfnamefont {M.}~\bibnamefont {Nicklas}}, \bibinfo {author} {\bibfnamefont
  {G.~M.}\ \bibnamefont {Pang}}, \bibinfo {author} {\bibfnamefont
  {L.}~\bibnamefont {Jiao}}, \bibinfo {author} {\bibfnamefont {W.~B.}\
  \bibnamefont {Jiang}}, \bibinfo {author} {\bibfnamefont {Y.}~\bibnamefont
  {Chen}}, \bibinfo {author} {\bibfnamefont {F.}~\bibnamefont {Steglich}},\
  and\ \bibinfo {author} {\bibfnamefont {H.~Q.}\ \bibnamefont {Yuan}},\
  }\bibfield  {title} {\bibinfo {title} {Two-gap superconductivity in
  {LaNiGa}$_{2}$ with nonunitary triplet pairing and even parity gap
  symmetry},\ }\href {https://doi.org/10.1103/PhysRevLett.117.027001}
  {\bibfield  {journal} {\bibinfo  {journal} {Phys. Rev. Lett.}\ }\textbf
  {\bibinfo {volume} {117}},\ \bibinfo {pages} {027001} (\bibinfo {year}
  {2016})}\BibitemShut {NoStop}%
\bibitem [{\citenamefont {Agterberg}\ \emph {et~al.}(1999)\citenamefont
  {Agterberg}, \citenamefont {Barzykin},\ and\ \citenamefont
  {Gor'kov}}]{Agterberg1999}%
  \BibitemOpen
  \bibfield  {author} {\bibinfo {author} {\bibfnamefont {D.~F.}\ \bibnamefont
  {Agterberg}}, \bibinfo {author} {\bibfnamefont {V.}~\bibnamefont
  {Barzykin}},\ and\ \bibinfo {author} {\bibfnamefont {L.~P.}\ \bibnamefont
  {Gor'kov}},\ }\bibfield  {title} {\bibinfo {title} {Conventional mechanisms
  for exotic superconductivity},\ }\href
  {https://doi.org/10.1103/PhysRevB.60.14868} {\bibfield  {journal} {\bibinfo
  {journal} {Phys. Rev. B}\ }\textbf {\bibinfo {volume} {60}},\ \bibinfo
  {pages} {14868} (\bibinfo {year} {1999})}\BibitemShut {NoStop}%
\bibitem [{\citenamefont {Ghosh}\ \emph {et~al.}(2018)\citenamefont {Ghosh},
  \citenamefont {Annett},\ and\ \citenamefont {Quintanilla}}]{Ghosh2018}%
  \BibitemOpen
  \bibfield  {author} {\bibinfo {author} {\bibfnamefont {S.}~\bibnamefont
  {Ghosh}}, \bibinfo {author} {\bibfnamefont {J.~F.}\ \bibnamefont {Annett}},\
  and\ \bibinfo {author} {\bibfnamefont {J.}~\bibnamefont {Quintanilla}},\
  }\bibfield  {title} {\bibinfo {title} {Time-reversal symmetry breaking in
  superconductors through loop {J}osephson-current order},\ }\href
  {https://arxiv.org/abs/1803.02618} {\bibfield  {journal} {\bibinfo  {journal}
  {arXiv preprint arXiv:1803.02618}\ } (\bibinfo {year} {2018})}\BibitemShut
  {NoStop}%
\bibitem [{\citenamefont {Xie}\ \emph {et~al.}(2020)\citenamefont {Xie},
  \citenamefont {Zhang}, \citenamefont {Shen}, \citenamefont {Jiang},
  \citenamefont {Pang}, \citenamefont {Shang}, \citenamefont {Gao},
  \citenamefont {Smidman},\ and\ \citenamefont {Yuan}}]{Xie2019}%
  \BibitemOpen
  \bibfield  {author} {\bibinfo {author} {\bibfnamefont {W.}~\bibnamefont
  {Xie}}, \bibinfo {author} {\bibfnamefont {P.~R.}\ \bibnamefont {Zhang}},
  \bibinfo {author} {\bibfnamefont {B.}~\bibnamefont {Shen}}, \bibinfo {author}
  {\bibfnamefont {W.~B.}\ \bibnamefont {Jiang}}, \bibinfo {author}
  {\bibfnamefont {G.~M.}\ \bibnamefont {Pang}}, \bibinfo {author}
  {\bibfnamefont {T.}~\bibnamefont {Shang}}, \bibinfo {author} {\bibfnamefont
  {C.}~\bibnamefont {Gao}}, \bibinfo {author} {\bibfnamefont {M.}~\bibnamefont
  {Smidman}},\ and\ \bibinfo {author} {\bibfnamefont {H.~Q.}\ \bibnamefont
  {Yuan}},\ }\bibfield  {title} {\bibinfo {title} {{Ca}{Pt}{As}: a new
  noncentrosymmetric superconductor},\ }\href
  {https://doi.org/10.1007/s11433-019-1488-5} {\bibfield  {journal} {\bibinfo
  {journal} {Sci. China-Phys. Mech. Astron.}\ }\textbf {\bibinfo {volume}
  {63}},\ \bibinfo {pages} {237412} (\bibinfo {year} {2020})}\BibitemShut
  {NoStop}%
\bibitem [{Sup()}]{Supple}%
  \BibitemOpen
  \href {https://doi.org/xxx} {}\bibinfo {note} {See the Supplemental Material
  at http://link.\-aps.\-org/\-sup\-ple\-men\-tal/xxx/PhysRevLett.xxxxx for
  details on the measurements of the upper critical field, the analyses of
  superfluid density, specific heat, and ZF-$\mu$SR data, and Refs.~[43--44,
  46--47] therein.}\BibitemShut {Stop}%
\bibitem [{\citenamefont {Tinkham}(1996)}]{tinkham1996}%
  \BibitemOpen
  \bibfield  {author} {\bibinfo {author} {\bibfnamefont {M.}~\bibnamefont
  {Tinkham}},\ }\href@noop {} {\emph {\bibinfo {title} {Introduction to
  Superconductivity}}},\ \bibinfo {edition} {2nd}\ ed.\ (\bibinfo  {publisher}
  {Dover Publications},\ \bibinfo {address} {Mineola, NY},\ \bibinfo {year}
  {1996})\BibitemShut {NoStop}%
\bibitem [{\citenamefont {Werthamer}\ \emph {et~al.}(1966)\citenamefont
  {Werthamer}, \citenamefont {Helfand},\ and\ \citenamefont
  {Hohenberg}}]{Werthamer1966}%
  \BibitemOpen
  \bibfield  {author} {\bibinfo {author} {\bibfnamefont {N.~R.}\ \bibnamefont
  {Werthamer}}, \bibinfo {author} {\bibfnamefont {E.}~\bibnamefont {Helfand}},\
  and\ \bibinfo {author} {\bibfnamefont {P.~C.}\ \bibnamefont {Hohenberg}},\
  }\bibfield  {title} {\bibinfo {title} {Temperature and purity dependence of
  the superconducting critical field, ${{H}}_{c2}$. {III}. {E}lectron spin and
  spin-orbit effects},\ }\href {https://doi.org/10.1103/PhysRev.147.295}
  {\bibfield  {journal} {\bibinfo  {journal} {Phys. Rev.}\ }\textbf {\bibinfo
  {volume} {147}},\ \bibinfo {pages} {295} (\bibinfo {year}
  {1966})}\BibitemShut {NoStop}%
\bibitem [{\citenamefont {Brandt}(2003)}]{Brandt2003}%
  \BibitemOpen
  \bibfield  {author} {\bibinfo {author} {\bibfnamefont {E.~H.}\ \bibnamefont
  {Brandt}},\ }\bibfield  {title} {\bibinfo {title} {Properties of the ideal
  {G}inzburg-{L}andau vortex lattice},\ }\href
  {https://doi.org/10.1103/PhysRevB.68.054506} {\bibfield  {journal} {\bibinfo
  {journal} {Phys. Rev. B}\ }\textbf {\bibinfo {volume} {68}},\ \bibinfo
  {pages} {054506} (\bibinfo {year} {2003})}\BibitemShut {NoStop}%
\bibitem [{\citenamefont {Padamsee}\ \emph {et~al.}(1973)\citenamefont
  {Padamsee}, \citenamefont {Neighbor},\ and\ \citenamefont
  {Shiffman}}]{Padamsee1973}%
  \BibitemOpen
  \bibfield  {author} {\bibinfo {author} {\bibfnamefont {H.}~\bibnamefont
  {Padamsee}}, \bibinfo {author} {\bibfnamefont {J.~E.}\ \bibnamefont
  {Neighbor}},\ and\ \bibinfo {author} {\bibfnamefont {C.~A.}\ \bibnamefont
  {Shiffman}},\ }\bibfield  {title} {\bibinfo {title} {Quasiparticle
  phenomenology for thermodynamics of strong-coupling superconductor},\ }\href
  {https://doi.org/10.1007/BF00654872} {\bibfield  {journal} {\bibinfo
  {journal} {J. Low Temp. Phys.}\ }\textbf {\bibinfo {volume} {12}},\ \bibinfo
  {pages} {387} (\bibinfo {year} {1973})}\BibitemShut {NoStop}%
\bibitem [{\citenamefont {Bouquet}\ \emph {et~al.}(2001)\citenamefont
  {Bouquet}, \citenamefont {Wang}, \citenamefont {Fisher}, \citenamefont
  {Hinks}, \citenamefont {Jorgensen}, \citenamefont {Junod},\ and\
  \citenamefont {Phillips}}]{Bouquet2001}%
  \BibitemOpen
  \bibfield  {author} {\bibinfo {author} {\bibfnamefont {F.}~\bibnamefont
  {Bouquet}}, \bibinfo {author} {\bibfnamefont {Y.}~\bibnamefont {Wang}},
  \bibinfo {author} {\bibfnamefont {R.~A.}\ \bibnamefont {Fisher}}, \bibinfo
  {author} {\bibfnamefont {D.~G.}\ \bibnamefont {Hinks}}, \bibinfo {author}
  {\bibfnamefont {J.~D.}\ \bibnamefont {Jorgensen}}, \bibinfo {author}
  {\bibfnamefont {A.}~\bibnamefont {Junod}},\ and\ \bibinfo {author}
  {\bibfnamefont {N.~E.}\ \bibnamefont {Phillips}},\ }\bibfield  {title}
  {\bibinfo {title} {Phenomenological two-gap model for the specific heat of
  {Mg}{B}$_{2}$},\ }\href {https://doi.org/10.1209/epl/i2001-00598-7}
  {\bibfield  {journal} {\bibinfo  {journal} {Europhys. Lett.}\ }\textbf
  {\bibinfo {volume} {56}},\ \bibinfo {pages} {856} (\bibinfo {year}
  {2001})}\BibitemShut {NoStop}%
\bibitem [{\citenamefont {Kubo}\ and\ \citenamefont {Toyabe}(1967)}]{Kubo1967}%
  \BibitemOpen
  \bibfield  {author} {\bibinfo {author} {\bibfnamefont {R.}~\bibnamefont
  {Kubo}}\ and\ \bibinfo {author} {\bibfnamefont {T.}~\bibnamefont {Toyabe}},\
  }\bibfield  {title} {\bibinfo {title} {A stochastic model for low field
  resonance and relaxation},\ }in\ \href@noop {} {\emph {\bibinfo {booktitle}
  {Magnetic Resonance and Relaxation}}},\ \bibinfo {editor} {edited by\
  \bibinfo {editor} {\bibfnamefont {R.}~\bibnamefont {Blinc}}}\ (\bibinfo
  {publisher} {North-Holland},\ \bibinfo {address} {Amsterdam},\ \bibinfo
  {year} {1967})\ pp.\ \bibinfo {pages} {810--823}\BibitemShut {NoStop}%
\bibitem [{\citenamefont {Yaouanc}\ and\ \citenamefont
  {de~R\'eotier}(2011)}]{Yaouanc2011}%
  \BibitemOpen
  \bibfield  {author} {\bibinfo {author} {\bibfnamefont {A.}~\bibnamefont
  {Yaouanc}}\ and\ \bibinfo {author} {\bibfnamefont {P.~D.}\ \bibnamefont
  {de~R\'eotier}},\ }\href@noop {} {\emph {\bibinfo {title} {Muon Spin
  Rotation, Relaxation, and Resonance: Applications to Condensed Matter}}}\
  (\bibinfo  {publisher} {Oxford University Press},\ \bibinfo {address}
  {Oxford},\ \bibinfo {year} {2011})\BibitemShut {NoStop}%
\bibitem [{\citenamefont {Bhattacharyya}\ \emph {et~al.}(2019)\citenamefont
  {Bhattacharyya}, \citenamefont {Adroja}, \citenamefont {Panda}, \citenamefont
  {Saha}, \citenamefont {Das}, \citenamefont {Machado}, \citenamefont
  {Cigarroa}, \citenamefont {Grant}, \citenamefont {Fisk}, \citenamefont
  {Hillier},\ and\ \citenamefont {Manfrinetti}}]{Bhattacharyya2019}%
  \BibitemOpen
  \bibfield  {author} {\bibinfo {author} {\bibfnamefont {A.}~\bibnamefont
  {Bhattacharyya}}, \bibinfo {author} {\bibfnamefont {D.~T.}\ \bibnamefont
  {Adroja}}, \bibinfo {author} {\bibfnamefont {K.}~\bibnamefont {Panda}},
  \bibinfo {author} {\bibfnamefont {S.}~\bibnamefont {Saha}}, \bibinfo {author}
  {\bibfnamefont {T.}~\bibnamefont {Das}}, \bibinfo {author} {\bibfnamefont
  {A.~J.~S.}\ \bibnamefont {Machado}}, \bibinfo {author} {\bibfnamefont
  {O.~V.}\ \bibnamefont {Cigarroa}}, \bibinfo {author} {\bibfnamefont {T.~W.}\
  \bibnamefont {Grant}}, \bibinfo {author} {\bibfnamefont {Z.}~\bibnamefont
  {Fisk}}, \bibinfo {author} {\bibfnamefont {A.~D.}\ \bibnamefont {Hillier}},\
  and\ \bibinfo {author} {\bibfnamefont {P.}~\bibnamefont {Manfrinetti}},\
  }\bibfield  {title} {\bibinfo {title} {Evidence of a nodal line in the
  superconducting gap symmetry of noncentrosymmetric {Th}{Co}{C}$_2$},\ }\href
  {https://doi.org/10.1103/PhysRevLett.122.147001} {\bibfield  {journal}
  {\bibinfo  {journal} {Phys. Rev. Lett.}\ }\textbf {\bibinfo {volume} {122}},\
  \bibinfo {pages} {147001} (\bibinfo {year} {2019})}\BibitemShut {NoStop}%
\bibitem [{\citenamefont {Samokhin}\ \emph {et~al.}(2004)\citenamefont
  {Samokhin}, \citenamefont {Zijlstra},\ and\ \citenamefont
  {Bose}}]{Samokhin2004}%
  \BibitemOpen
  \bibfield  {author} {\bibinfo {author} {\bibfnamefont {K.~V.}\ \bibnamefont
  {Samokhin}}, \bibinfo {author} {\bibfnamefont {E.~S.}\ \bibnamefont
  {Zijlstra}},\ and\ \bibinfo {author} {\bibfnamefont {S.~K.}\ \bibnamefont
  {Bose}},\ }\bibfield  {title} {\bibinfo {title} {{CePt$_3$Si}: {An}
  unconventional superconductor without inversion center},\ }\href
  {https://doi.org/10.1103/PhysRevB.69.094514} {\bibfield  {journal} {\bibinfo
  {journal} {Phys. Rev. B}\ }\textbf {\bibinfo {volume} {69}},\ \bibinfo
  {pages} {094514} (\bibinfo {year} {2004})}\BibitemShut {NoStop}%
\bibitem [{\citenamefont {Lee}\ and\ \citenamefont {Pickett}(2005)}]{Lee2005}%
  \BibitemOpen
  \bibfield  {author} {\bibinfo {author} {\bibfnamefont {K.-W.}\ \bibnamefont
  {Lee}}\ and\ \bibinfo {author} {\bibfnamefont {W.~E.}\ \bibnamefont
  {Pickett}},\ }\bibfield  {title} {\bibinfo {title} {Crystal symmetry,
  electron-phonon coupling, and superconducting tendencies in
  {Li}$_2${Pd}$_3${B} and {Li}$_2${Pt}$_3${B}},\ }\href
  {https://doi.org/10.1103/PhysRevB.72.174505} {\bibfield  {journal} {\bibinfo
  {journal} {Phys. Rev. B}\ }\textbf {\bibinfo {volume} {72}},\ \bibinfo
  {pages} {174505} (\bibinfo {year} {2005})}\BibitemShut {NoStop}%
\bibitem [{\citenamefont {Frigeri}\ \emph {et~al.}(2004)\citenamefont
  {Frigeri}, \citenamefont {Agterberg}, \citenamefont {Koga},\ and\
  \citenamefont {Sigrist}}]{Frigeri2004}%
  \BibitemOpen
  \bibfield  {author} {\bibinfo {author} {\bibfnamefont {P.~A.}\ \bibnamefont
  {Frigeri}}, \bibinfo {author} {\bibfnamefont {D.~F.}\ \bibnamefont
  {Agterberg}}, \bibinfo {author} {\bibfnamefont {A.}~\bibnamefont {Koga}},\
  and\ \bibinfo {author} {\bibfnamefont {M.}~\bibnamefont {Sigrist}},\
  }\bibfield  {title} {\bibinfo {title} {Superconductivity without inversion
  symmetry: {MnSi versus
  ${\mathrm{C}\mathrm{e}\mathrm{P}\mathrm{t}}_{3}\mathrm{S}\mathrm{i}$}},\
  }\href {https://doi.org/10.1103/PhysRevLett.92.097001} {\bibfield  {journal}
  {\bibinfo  {journal} {Phys. Rev. Lett.}\ }\textbf {\bibinfo {volume} {92}},\
  \bibinfo {pages} {097001} (\bibinfo {year} {2004})}\BibitemShut {NoStop}%
\bibitem [{\citenamefont {Schnyder}\ and\ \citenamefont
  {Brydon}(2015)}]{Schnyder2015}%
  \BibitemOpen
  \bibfield  {author} {\bibinfo {author} {\bibfnamefont {A.~P.}\ \bibnamefont
  {Schnyder}}\ and\ \bibinfo {author} {\bibfnamefont {P.~M.~R.}\ \bibnamefont
  {Brydon}},\ }\bibfield  {title} {\bibinfo {title} {Topological surface states
  in nodal superconductors},\ }\href
  {https://doi.org/10.1088/0953-8984/27/24/243201} {\bibfield  {journal}
  {\bibinfo  {journal} {J. Phys.: Condens. Matter}\ }\textbf {\bibinfo {volume}
  {27}},\ \bibinfo {pages} {243201} (\bibinfo {year} {2015})}\BibitemShut
  {NoStop}%
\bibitem [{\citenamefont {Sato}\ and\ \citenamefont {Ando}(2017)}]{Sato2017}%
  \BibitemOpen
  \bibfield  {author} {\bibinfo {author} {\bibfnamefont {M.}~\bibnamefont
  {Sato}}\ and\ \bibinfo {author} {\bibfnamefont {Y.}~\bibnamefont {Ando}},\
  }\bibfield  {title} {\bibinfo {title} {Topological superconductors: a
  review},\ }\href {https://doi.org/10.1088/1361-6633/aa6ac7} {\bibfield
  {journal} {\bibinfo  {journal} {Rep. Prog. Phys.}\ }\textbf {\bibinfo
  {volume} {80}},\ \bibinfo {pages} {076501} (\bibinfo {year}
  {2017})}\BibitemShut {NoStop}%
\end{thebibliography}%

\end{document}